\documentclass[12pt,emcite]{article}
\usepackage{amsmath,amssymb,amsthm}
\usepackage{color}
\usepackage{graphicx}
\usepackage{enumerate}
\usepackage{harvard}

\theoremstyle{plain}
\newtheorem{theorem}{Theorem}
\newtheorem{corollary}{Corollary}
\newtheorem{lemma}{Lemma}
\newtheorem{proposition}{Proposition}

\theoremstyle{definition}
\newtheorem{definition}{Definition}
\newtheorem{assumption}{Assumption}

\renewcommand{\cite}{\citeasnoun}
\begin{document}

\begin{titlepage}

\title{{\bf Uncertainty in Mechanism Design}\thanks{
We thank David Ahn, Ben Polak, Debraj Ray, and Ennio Stacchetti for helpful discussions and comments. Shannon thanks the NSF for support under grants SES 0721145 and SES 1227707, and the Center for Advanced Study in the Behavioral Sciences at Stanford University.} 
}

\author{
Giuseppe Lopomo\thanks{%
The Fuqua School of Business, Duke University; \tt{glopomo@duke.edu}} 
\\ Fuqua School of Business\\ Duke University
\and
Luca Rigotti\thanks{%
University of Pittsburgh; \tt{luca@pitt.edu}} 
\\ Department of Economics\\ University of Pittsburgh
\and 
Chris Shannon\thanks{%
UC Berkeley; \tt{cshannon@econ.berkeley.edu}} 
\\ Department of Economics\\ U.C. Berkeley 
}

\date{December 2020}

\maketitle

\begin{abstract}
This paper studies the design of mechanisms that are robust to
misspecification. We introduce a novel notion of robustness that connects a
variety of disparate approaches and study its implications in a wide class
of mechanism design problems. This notion is quantifiable, allowing us to
formalize and answer comparative statics questions relating the nature and
degree of misspecification to sharp predictions regarding features of
feasible mechanisms. This notion also has a behavioral foundation which reflects the perception of ambiguity, thus allowing the degree of misspecification to emerge endogenously. In a number of standard settings, robustness to
arbitrarily small amounts of misspecification generates a discontinuity in
the set of feasible mechanisms and uniquely selects simple, ex post
incentive compatible mechanisms such as second-price auctions. Robustness
also sheds light on the value of private information and the prevalence of
full or virtual surplus extraction.
\end{abstract}

\vspace{3mm}


\thispagestyle{empty}

\end{titlepage}

\newpage

\section{Introduction}

This paper studies the design of mechanisms that are robust to
misspecification. We introduce a novel notion of robustness that connects a
variety of disparate approaches and study its implications in a wide class
of mechanism design problems. This notion is quantifiable, allowing us to
formalize and answer comparative statics questions relating the nature and
degree of misspecification to sharp predictions regarding features of
feasible mechanisms. We show that in a number of standard settings robustness to
arbitrarily small amounts of misspecification generates a discontinuity in
the set of feasible mechanisms and uniquely selects simple, ex post
incentive compatible mechanisms such as second-price auctions. This notion of robustness
also sheds light on the value of private information and the prevalence of
full or virtual surplus extraction.

We model misspecification through imprecision in agents' beliefs. We
consider a generalization of standard Bayesian mechanism design problems in
which each agent is associated with a set of distributions over the
underlying state space, rather than a unique distribution. As in standard
settings with correlated beliefs, this set can depend on the agent's type,
which is his private information. We model robustness to this
misspecification by focusing on mechanisms requiring interim incentive
compatibility to hold in expectation for each distribution in this set.

Our notion of robustness can be motivated in several distinct ways,
each with a rich recent history. One argument follows the ``Wilson doctrine'',
calling for reworking models of strategic
interactions with incomplete information to reduce the reliance on common prior assumptions and precise information about beliefs and higher-order beliefs, or other fine details, and to develop theories that are
``detail-free'' instead. Much work in this vein
adopts alternative ``belief-free'' solution
concepts such as ex post equilibrium. Another growing body of work instead
defines robustness more primitively and then studies its implications in a
variety of settings. 

A second motivation for our notion of robustness instead starts from the ideas of \cite{Knight21} and \cite{Ellsberg}. In
many important economic settings, uncertainty is not described by a precise
probability distribution over the relevant outcomes, and agents' choices
cannot be rationalized using a precise probability distribution in any
probabilistically sophisticated way. A large literature developing models of
individual decision making under ambiguity has grown from these foundational ideas. Other work builds on these ideas by exploring  implications of ambiguity in individual choice problems and equilibrium. Allowing for
ambiguity is particularly natural and important in strategic interactions
with incomplete information: even agents with precise beliefs about some primitive parameters of uncertainty might be  ambiguous about opponents' beliefs, or about whether they hold precise beliefs.\footnote{\cite{Ahn07} proves the existence of an analogue of the Mertens-Zamir universal type space for hierarchies of ambiguous sets of beliefs. As he shows, ambiguity at any level of the belief hierarchy corresponds to a set of beliefs over types.}

Our results can be viewed as studying the extent to which
ambiguity influences outcomes in simple incomplete information games. Each type is associated with a set of probability distributions in our
model, and ambiguity can be understood as in Bewley (1986) via incomplete preferences,
or as objective uncertainty, as in \cite{Ghirardato-Maccheroni-Marinacci} and \cite{Gilboa-Maccheroni-Marinacci-Schmeidler}. In this sense, our
notion of robustness has a behavioral interpretation. An advantage of this
approach is that the set of beliefs emerges endogenously from preferences and the size of this set has a clear interpretation as the amount of ambiguity perceived. Moreover, this
framework gives decision-theoretic foundations to questions of robust
mechanism design.

The first main result provides conditions under which mechanisms satisfying our notion of robustness must have simple structure. We identify a notion of interim incentive compatibility robust to misspecification, which we call  robust incentive
compatibility, and show that under certain conditions this implies an ex post envelope condition requiring that differences in ex post utilities across types are pinned down by the allocation rule. These
conditions include standard regularity assumptions, such as a continuum of
types and smoothness of the utility function. They also include novel
restrictions on the richness of beliefs through a condition we
call fully overlapping beliefs. In particular this includes the focal
special case in which each type is associated with an arbitrarily small
neighborhood around a fixed belief. In many settings the restriction of feasible mechanisms to those satisfying this ex post envelope condition will severely
limit the scope of a mechanism designer. For example, we show that if in
addition a mechanism satisfies a general monotonicity condition, then it must be ex post incentive compatible. To illustrate, we consider the particular case of quasilinear auctions. Here we show that for auction mechanisms that are monotone in types, if beliefs are
fully overlapping then the only robust incentive compatible auctions are
those that are ex post incentive compatible. 

Our second set of results considers a robust version of the 
surplus extraction problem. In standard Bayesian models, the designer can typically extract all, or virtually all, information rents  whenever agents' beliefs are correlated with their private information, following from the foundational work of Cr{\'e}mer and McLean (1985, 1988) and \cite{McAfeeReny92}.\nocite{CremerMcLean85}\nocite{CremerMcLean88}
This has been a central puzzle in mechanism design, and has motivated significant attention to developing foundations for mechanisms less sensitive to fine  details of the environment, particularly agents' beliefs. To address this, we first give a robust generalization of the results of Cr{\'e}mer and McLean (1985, 1988) and \cite{McAfeeReny92}.  We show that the designer can achieve a robust version of virtual extraction whenever agents' beliefs satisfy a natural set-valued analogue of the convex independence conditions of Cr{\'e}mer-McLean and McAfee-Reny. Virtual extraction frequently fails in our setting, in contrast with the standard Bayesian model in which full or virtual extraction is generically possible in many contexts. We then study limits on the designer's ability to extract surplus. When virtual extraction fails, we show that the designer can be restricted to simple mechanisms. In particular, when beliefs are fully overlapping, robust incentive compatibility limits the designer to offering a single contract, and additional natural conditions can make a deterministic contract optimal for the designer.

Our work is related to the recent literature on robust mechanism design. Central in this literature is \cite{Bergemann-Morris05}%
. They define robustness in a mechanism design context as implementation in
the universal type space of \cite{Mertens-Zamir}. This provides an elegant and
powerful way to formalize the requirement that mechanisms should be robust
to fine details of higher-order beliefs. They show that implementation in
the universal type space is equivalent to ex post implementation. Our results can be interpreted as generalizing the \cite{Bergemann-Morris05} notion of robustness while nesting it
within a class that ranges from the standard Bayesian model at one extreme,
in which there is no misspecification and each type of each agent is
associated with a unique belief, to the case of complete misspecification at
the other extreme, in which every type of every agent is associated with the
set of all possible beliefs.
By
nesting these extremes within a larger class, we can study the implications of
robustness in important intermediate cases. In particular, this allows us to
consider the important special case of arbitrarily small amounts of
misspecification. This also allows us to identify conditions on the map from
types to beliefs under which robustness requires the use of simple
mechanisms. In contrast, the arguments of \cite{Bergemann-Morris05} are tailored
to their assumption of complete misspecification and cannot shed light on
the implications of intermediate cases or on the robustness of predictions in the standard Bayesian model to arbitrarily small amounts of misspecification.

Our results are also connected to other recent work on robustness in mechanism design problems. 
  \cite{Jehiel-Meyer-ter-Vehn-Moldovanu-12} adopt the model of robustness we introduce in this paper, and show that ex post implementation is generically impossible in the leading example of epsilon-contamination we introduce in the next section. \cite{Chiesa-Micali-Zhu-15} adapt our model to a setting with a finite set of types, and focus on the performance of Vickrey mechanisms in either dominant or undominated strategies. A number of papers consider distributionally robust versions of standard problems, in which the designer does not know all relevant distributions but might have some partial information, such as that distributions belong to a set with certain properties, bounds on the supports, or restrictions on means or other moments. Much of this work studies optimal mechanisms under worst case maxmin revenue guarantees or other revenue or welfare goals, including \cite{Auster}, \cite{Bergemann-Schlag-a}, \cite{Bergemann-Schlag-b}, \cite{Carrasco-Luz-Kos-Messner-Monteiro}, \cite{Hu-Haghpanah-Hartline-Kleinberg}, \cite{Kocyigit-Iyengar-Kuhn-Wiesemann-20}, \cite{Neeman}, and \cite{Suzdaltsev}. Similarly, Oll{\' a}r and Penta (2017, 2019) study full and partial implementation under moment conditions on the set of possible distributions.  \nocite{Ollar-Penta-17}  \nocite{Ollar-Penta-19}

Our paper is also related to a growing literature on mechanism design with ambiguity-averse agents. \cite{Bodoh-Creed-12}, \cite{Bose-Ozdenoren-Pape-06}, and \cite{Wolitzky-16} extend standard mechanism design problems to allow for maxmin expected utility agents, while \cite{Bose-Renou-14} and \cite{DeTillio-Kos-Messner-17} allow for maxmin expected utility agents facing ambiguity about aspects of the mechanism. \cite{Bose-Daripa-09} consider an independent private values setting in which bidders have maxmin expected utility preferences, with beliefs modeled using epsilon-contamination. They show that the seller can extract almost all of the surplus by using a dynamic mechanism which is a modified Dutch auction.

The paper proceeds as follows. The next section presents our main result in
a single-agent, quasilinear, mechanism design setting. Section \ref%
{section:Auctions} illustrates how this result extends to a multi-agent
auction model, while section \ref{section:GeneralModel} illustrates how it
extends to the case of more general utility functions. Section \ref%
{section:Preferences} presents the decision-theoretic model needed to
interpret our notion of robustness in terms of ambiguity. Section \ref%
{section:FullExtraction} studies information rents and  surplus
extraction. Section 7 concludes. The appendix collects material and proofs omitted from the main body of the paper. 

\section{\label{section:Quasilinear} Quasilinear Model}

In this section, we develop the main ideas and results in the context of a
simple quasilinear setting. This highlights the role of robustness as we define it and of the
novel conditions we introduce in a simplified standard environment. We present the basic setup first in section
2.1, along with notation and assumptions used throughout the paper. Then we turn to the main conditions on beliefs in section 2.2; these
will also be used throughout the paper. We give the main result for the
quasilinear model in section 2.3.

\subsection{Setup}

Consider a seller who designs a mechanism to transfer an indivisible object to
a buyer. The buyer is privately informed about his valuation for the object,
denoted $t\in T =[0,1]$. The seller chooses a direct mechanism that
depends on the buyer's reported type and an uncertain state of the world $s$
drawn from a set $S$. The realized state is publicly verifiable.  Throughout we assume that the set of states $S$ is a compact metric space. 

We use the following standard notation. For a compact metric space $B$, $C(B)$ is the space of continuous real-valued functions on $B$, and $\Delta(B)$ is the space of Borel probability measures on $B$. We take $\Delta(B)$ to be endowed with the weak$^*$ topology unless otherwise stated. For $b\in B$, $\delta_b\in \Delta(B)$ denotes the Dirac measure concentrated on $b$.  For a subset $A$ of a topological vector space $X$, $\mbox{co} (A)$ denotes the convex hull of $A$, and $\overline{\mbox{co}} (A)$ denotes the closed convex hull of $A$.

Throughout the paper we restrict attention to direct mechanisms. In addition, we assume agents do not randomize  their reports in these mechanisms. Both restrictions are without loss of generality for our results. In the appendix, we show that a version of the revelation principle holds in our setting. We discuss the issue of randomizing reports in more detail below following the definition of robust incentive compatibility.  

In this section, we consider a simple quasilinear setting; in section \ref{section:GeneralModel} we show that this restriction is not important for our main conclusions. Here the seller chooses a direct mechanism $(q,p)$ consisting of an allocation rule $q$ and a payment function $p$. The allocation rule $q:T\times S\rightarrow [0,1]$
specifies the probability the buyer gets the object as a function of his
report and the realized state. The payment scheme $p:T\times S\rightarrow 
\mathbf{R}$ specifies how much the buyer pays as a function of his report
and the realized state. We assume that $q(t, \cdot)$ and $p(t,\cdot)$ are Borel measurable for each $t\in T$. 

The buyer has quasilinear utility over valuations and payments. When he
reports $\theta $ while his true type is $t$, his (expected) utility in state $s$ is 
\begin{equation*}
tq(\theta ,s)-p(\theta ,s)
\end{equation*}
Given $\pi \in \Delta (S)$, the buyer's expected utility from reporting
$\theta $ while his true type is $t$ is 
\begin{equation*}
E_{\pi}\left[ tq(\theta ,s)-p(\theta ,s)\right] 
\end{equation*}
This expectation is taken at the interim stage, when the buyer knows his type but does not know the realized state. 

This is a standard monopolistic screening model, with the complication that the mechanism can depend on the exogenous state. This dependence allows us to specialize the model in different directions. For example,
we can easily extend this framework to the case of many buyers by adding
additional agents and modifying the state space to include the set of their possible type profiles. In section %
\ref{section:Auctions} we illustrate this extension in the setting of a
quasilinear auction model. We can also consider questions of surplus
extraction with correlated beliefs, as we do in section \ref%
{section:FullExtraction}.

Next we present three nested notions of incentive compatibility. The first
two are standard while the third is novel and designed to reflect robustness
to misspecification of agents' beliefs. The first requires incentive
compatibility to hold after uncertainty about the state of the world is
realized; the second and third, instead, apply at the interim stage when the
agent knows his type but does not know which state has occurred. 

We start with ex post incentive compatibility, in which reporting the true type is
the agent's preferred strategy regardless of which state obtains.

\begin{definition}
\label{definition:EPICql}A mechanism $(q,p)$ is \emph{ex post incentive
compatible} if for each $t, \theta \in T$, 
\begin{equation*}
tq(t,s)-p(t,s)\geq  tq(\theta ,s)-p(\theta ,s)  \ \ \forall s\in S
\end{equation*}
\end{definition}

The second notion is Bayesian or interim incentive compatibility, which requires that reporting the true
type is the agent's preferred strategy in expectation.

\begin{definition}
\label{definition:BICql}A mechanism $(q,p)$ is \emph{interim incentive
compatible} given $\{ \pi(t) \in \Delta(S) : t\in T\}$ if for each $t, \theta \in T$,
\[
E_{\pi (t)}[tq(t,s)-p(t,s)]\geq E_{\pi (t)}\left[ tq(\theta
,s)-p(\theta ,s)\right] 
\]
\end{definition}

These notions are clearly nested: ex post incentive compatibility implies interim 
incentive compatibility for any $\{ \pi(t)\in \Delta(S): t\in T\}$.

We seek an intermediate notion of incentive compatibility that reflects robustness to 
misspecification of beliefs $\{\pi (t) \in \Delta(S):t\in T\}$. To that end, for each type $t\in T$, fix a set $\Pi
(t)\subseteq \Delta (S)$. This set reflects the range of possible
misspecification of beliefs for type $t$. To ensure that incentive
compatibility is robust to this misspecification, interim incentive compatibility is then required to hold for every element of $\Pi (t)$.

\begin{definition}
\label{definition:OICql}A mechanism $(q,p)$ is \emph{robust incentive
compatible} for beliefs $\{ \Pi(t) \subseteq \Delta(S): t\in T\}$ if for each $t, \theta \in T$, 
\begin{equation*}
E_{\pi }[tq(t,s)-p(t,s)]\geq E_{\pi }\left[  tq(\theta
,s)-p(\theta ,s) \right]\ \ \forall \pi \in \Pi(t)
\end{equation*}
\end{definition}

The set $\Pi (t)$ plays two roles in the notion of robust incentive
compatibility. One is akin to beliefs in the standard notion of interim incentive compatibility. The second is to represent robustness against errors in the
specification of $\pi(t)$. For now, we take the collection $\{\Pi (t) \subseteq \Delta(S):t\in T\}$ to be given exogenously
and focus on the relationship between properties of these sets and the
nature of mechanisms that satisfy this notion of robustness. In section \ref%
{section:Preferences} we discuss a decision-theoretic model in which these sets 
emerge endogenously from preferences. We put no restrictions on the sets $\Pi(t)$ in our basic setup, although we note that it would be without loss of generality to assume that $\Pi(t)$ is closed and convex for each $t\in T$. That is, a mechanism $(q,p)$ is robust incentive compatible for beliefs $\{\Pi(t) \subseteq \Delta(S): t\in T\}$ if and only if it is robust incentive compatible for beliefs $\{ \overline{\mbox{co}}(\Pi(t)) : t\in T\}$. A similar observation holds in the extension to multiple agents we consider in section 3, and in the general model we consider in section 4. We return to this observation at several points below.  

The collection $\{\Pi (t) \subseteq \Delta(S):t\in T\}$ could be
constructed to reflect additional natural characteristics. For example, beliefs can be independent of type by assuming that there is a fixed set $\Pi \subseteq \Delta(S)$ with $\Pi(t) = \Pi$ for each type $t \in T$. Throughout we refer to such beliefs as {\it independent}. Similarly, these sets could be derived from a common set of priors by using Bayesian updating belief by belief, or could be members of a given parametric family. Features like common priors or independence are consistent with this framework, but are not required for our main results. 

As we mentioned above, we restrict attention throughout to direct mechanisms in which agents do not randomize their reports, and this is essentially without loss of generality in our setting. For example, note that if $(q, p)$ is robust incentive compatible, then it remains robust incentive compatible if the agent randomizes reports, under requisite additional regularity conditions. That is, suppose $q$ and $p$ are jointly Borel measurable and $p$ is bounded. Given $t\in T$ and $\pi\in \Pi(t)$, if $\sigma \in \Delta(T)$ is a randomized report, then 
\[
E_\pi[t q(t,s) - p(t,s)] \geq E_\sigma [ E_\pi[ tq(\theta, s) - p(\theta, s)]] = E_\pi[ E_\sigma[ tq(\theta, s) - p(\theta, s)]]
\]
where the first inequality follows from robust incentive compatibility.

As the example below illustrates, a natural way to model misspecification is to consider $\varepsilon$ mixtures with a fixed reference belief $\pi(t) \in \Delta(S)$. This will be the leading example for the paper. Here we use this example first to illustrate the meaning of robust incentive compatibility and the sense in which this notion gives a robustness check on the Bayesian model. 

\noindent {\bf Example 1: The $\varepsilon$-contamination model. } 
Fix $\varepsilon \in (0,1]$ and $\pi(t)\in \Delta (S)$ for each $t\in T$.
Let 
\begin{equation*}
\Pi _{\varepsilon }(t):=\{(1-\varepsilon )\pi(t)+\varepsilon \pi : \pi \in \Delta (S)\}
\end{equation*}%
This set consists of all measures that are mixtures of the reference measure $\pi(t)$ with another measure $\pi \in \Delta(S)$, with weights $1-\varepsilon$ and $\varepsilon$.

To illustrate the notion of robust incentive compatibility, suppose $\varepsilon \in (0,1)$. In this case, a mechanism is robust
incentive compatible if for each $t, \theta \in T$, 
\begin{multline*}
E_{\pi(t)}[tq(t,s)-p(t,s)] \geq E_{\pi(t)}\left[  tq(\theta ,s)-p(\theta ,s)\right]  \\
 +\frac{\varepsilon }{1-\varepsilon }E_{\pi}\left[
 tq(\theta ,s)-p(\theta ,s) -(tq(t,s)-p(t,s))\right] \
\ \forall \pi\in \Delta (S) 
\end{multline*}%
Equivalently, this requires that for each $t, \theta \in T$, 
\begin{multline*}
E_{\pi(t)}[tq(t,s)-p(t,s)] \geq E_{\pi(t)}\left[  tq(\theta ,s)-p(\theta ,s)\right] \\
 +\frac{\varepsilon }{1-\varepsilon }\sup_{\pi\in \Delta
(S)}E_{\pi}\left[tq(\theta ,s)-p(\theta ,s) 
-(tq(t,s)-p(t,s))\right] 
\end{multline*}%
In this example, robust incentive compatibility requires first that standard
interim incentive compatibility holds for the reference beliefs $\{ \pi(t) \in \Delta(S) :t\in T\}$, and in addition that it holds robustly for each type $t$ and report $\theta \in  T$,
with error term given by 
\begin{eqnarray*}
\frac{\varepsilon}{1-\varepsilon} \sup_{\pi \in \Delta(S)} E_{\pi} \left[ tq(\theta
,s)-p(\theta ,s) - (tq(t,s)-p(t,s) ) \right] 
\end{eqnarray*}

This is a focal example in either of the main motivations for our work. From the perspective of robustness to misspecification of beliefs and the ``Wilson doctrine'', this example is the natural case of $\varepsilon$-misspecification of a baseline prior $\pi(t)$. In this sense, this example gives a natural notion of local robustness.\footnote{For example, Jehiel, Meyer-ter-Vehn, Molodovanu (2012) adopt this simplified version of our model as a notion of local robustness and study the generic possibility of locally robust implementation in this sense.} From the perspective of robustness to ambiguity, this example is the standard $\varepsilon$-contamination model of multiple priors. As the parameter $\varepsilon$ ranges from 0 to 1, this example traces the spectrum from complete specification to complete misspecification or complete ambiguity.\hfill $\diamondsuit$

The three notions of incentive compatibility above are nested, with robust
incentive compatibility in the middle: an ex post incentive
compatible mechanism is also robust and interim incentive compatible,
while a robust incentive compatible mechanism is interim incentive
compatible (provided $\pi (t)\in \Pi (t)$ for each $t$). 

Our main results
give a characterization of robust incentive compatible mechanisms and
explore how these mechanisms depend on properties of beliefs $\{\Pi
(t) \subseteq \Delta(S) :t\in T\}$. Two results along these lines are immediate. When $\Pi (t)$
is a singleton for each $t$, by definition robust incentive compatibility
reduces to interim incentive compatibility. This corresponds to one
extreme, the case of no misspecification of beliefs. At the other extreme,
if $\Pi (t)=\Delta (S)$ for each $t$, then robust incentive compatibility
is equivalent to ex post incentive compatibility. This corresponds instead
to the case of complete misspecification. We record this simple observation
below.

\begin{lemma}
\label{lemma: full ignorance} If $\Pi (t)=\Delta (S)$ for each $t\in T$,
then a mechanism $(q,p)$ is robust incentive compatible for beliefs $\{\Pi(t): t\in T\}$  if and only if it
is ex post incentive compatible.
\end{lemma}

\begin{proof}
Ex post incentive compatibility implies robust incentive
compatibility for any beliefs $\{ \Pi(t) \subseteq \Delta(S): t\in T\}$, so we need to prove only
the converse. Then let $\Pi(t) = \Delta(S)$ for each $t\in T$,
and let $(q,p)$ be given. If $(q,p)$ is not ex post incentive compatible, then there exists
a state $s$, a type $t$, and a report $\theta \in T$ such that 
\begin{equation*}
tq(t,s)-p(t,s)<tq(\theta ,s)-p(\theta ,s)
\end{equation*}%
Let $\delta_s$ denote the Dirac measure assigning probability one to this state $s$%
. Then
\[
tq(t,s)-p(t,s) = E_{\delta_s}\left[tq(t,s)-p(t,s)\right] < E_{\delta_s}\left[tq(\theta ,s)-p(\theta ,s) \right] = tq(\theta ,s)-p(\theta ,s)
\]
Since $\delta_s\in \Pi( t) =\Delta (S)$, this implies $(q,p)$ is not a robust incentive
compatible mechanism for beliefs $\{ \Pi(t) : t\in T\}$. Thus if $(q,p)$ is robust incentive compatible for these beliefs, then it is ex post incentive compatible. 
\end{proof}

This observation is not new, as a related result appears in Ledyard (1978, 1979).\nocite{Ledyard78}\nocite{Ledyard79} For intermediate cases in which $\Pi(t)$ is neither a singleton nor the entire set $\Delta(S)$, the set of robust incentive compatible mechanisms will lie between the set of ex post and interim incentive compatible mechanisms. A central goal of the paper is to explore this intermediate case. Although it might be natural to conjecture that for belief sets close to singletons, and thus for models close to the standard Bayesian model, robust  mechanisms will have features close to those of standard interim incentive compatible mechanisms, we show that even when belief sets are arbitrarily small, robust incentive compatibility can have important ex post restrictions.

\subsection{Fully Overlapping Beliefs}

Our main results in the next three sections rely on two novel conditions on the richness of the set of beliefs,
one that applies to a single type, and the other relating beliefs across nearby types.
We develop these conditions next.

The first notion captures  richness of beliefs in a given set $\Pi \subseteq \Delta (S)$.

\begin{definition}
\label{definition:FullDimension} A set $\Pi \subseteq \Delta (S)$ has \emph{%
full dimension} if, given any Borel measurable function $g:S\rightarrow \mathbf{R}$, 
\begin{equation*}
E_\pi [ g(s) ] =0\ \ \forall \pi \in \Pi \Rightarrow g=0
\end{equation*}
\end{definition}

The $\varepsilon $-contamination model of Example 1 provides a simple illustration. Here 
the set $\Pi _{\varepsilon }(t)$ has full dimension
for any $\varepsilon \in (0,1]$. To see this, let $\varepsilon \in (0,1]$, and suppose $g:S\to {\bf R}$ is Borel measurable and $E_\pi [g(s)]  =0$ for all $\pi \in \Pi_{\varepsilon}(t)$. Since $\pi(t)\in \Pi_\varepsilon(t)$, this implies $E_{\pi(t)} [ g(s) ] =0$. Now let $\pi\in \Delta(S)$. 
By definition, $(1-\varepsilon)\pi(t) + \varepsilon \pi \in \Pi_\varepsilon(t)$, so 
\[
(1-\varepsilon)E_{\pi(t)} [ g(s) ] + \varepsilon E_\pi [ g(s) ] = \varepsilon E_\pi [ g(s) ] =0
\]
Since $\varepsilon >0$ and $\pi\in \Delta(S)$ were arbitrary, this implies $E_\pi [ g(s) ] = 0$ for all $\pi\in \Delta(S)$, which implies $g=0$. 

When the state space $S$ is finite, a set $\Pi \subseteq \Delta(S) $
has full dimension provided it contains a set of linearly independent
elements. More generally, $\Pi$ has full
dimension whenever its algebraic interior in $\Delta (S)$ is non-empty,
where the algebraic interior of $\Pi $ is given by 
\begin{equation*}
\text{alg-int }\Pi :=\{\pi \in \Pi :\forall \tilde{\pi}\in \Delta (S)\  \exists \delta \in \left( 0,1\right] \text{ such that }(1-\delta )\pi
+\delta \tilde{\pi}\in \Pi \}
\end{equation*}%
A stronger sufficient condition, equivalent when $S$ is finite, is that $\Pi 
$ has non-empty relative interior (relative to the affine hull of $\Delta(S)$). For $\Pi\subseteq \Delta(S)$, we let $\mbox{rint } \Pi$ denote the relative interior of $\Pi$. 

The second notion requires that nearby types share a sufficiently rich set of beliefs,
where richness is again in the sense of full dimension.

\begin{definition}
\label{definition:fullyoverlappingbeliefs} Beliefs $\{ \Pi(t) \subseteq \Delta(S): t\in
T\} $ are \emph{fully overlapping} if for each $t\in T$ there exists a
neighborhood $N(t)$ of $t$ such that $\bigcap\limits_{t^{\prime }\in
N(t)}\Pi (t^{\prime })$ has full dimension.
\end{definition}

If beliefs are independent, so all types $t\in T$ share a common set of beliefs $\Pi(t) = \Pi$ for some set $\Pi\subseteq \Delta(S)$, then beliefs are fully overlapping if and only if the set $\Pi$ has full dimension. 
More generally, when beliefs might vary with type, fully overlapping beliefs require that locally types share some set of beliefs, and that these locally common beliefs are a set of full dimension. 

We give two results next that provide sufficient conditions for beliefs to be fully overlapping, one for general beliefs and one for our leading example of $\varepsilon$-contamination. The first result comes from considering properties of the correspondence $\Pi :T\rightarrow 2^{\Delta(S)}$ that maps each type $t$ into the corresponding set $\Pi(t)$. As we show below, when the relative interior of $\Pi(t)$ is nonempty for each $t$, beliefs are fully overlapping under continuity conditions on this correspondence.

\begin{theorem}
Suppose for every $t\in T$, $\Pi(t)\subseteq \Delta(S)$ is convex and $\mbox{\rm rint } \Pi(t) \not= \emptyset$. 
\begin{itemize}
  \item[(i)] Let $\mbox{\rm rint } \Pi:T\to 2^{\Delta(S)}$ be the correspondence such that $(\mbox{\rm rint } \Pi) (t) = \mbox{\rm rint }\Pi(t)$ for each $t\in T$. If the graph of $\mbox{\rm rint } \Pi$ is relatively open, then beliefs are fully overlapping.
  \item[(ii)] If $S$ is finite and $\Pi:T\to 2^{\Delta(S)}$ is lower hemicontinuous, then beliefs are fully overlapping.   
\end{itemize}
\end{theorem}
\begin{proof}
For (i), let $t_0\in T$ and $\pi\in \mbox{rint } \Pi(t_0)$. Since the graph of $\mbox{rint } \Pi$ is relatively open, there exists a relatively open set $U\subseteq \Delta(S)$ with $\pi \in U$ and a neighborhood $N(t_0)$ of $t_0$ such  that $U \subseteq \mbox{rint } \Pi(t)$ for all $t\in N(t_0)$. 
Then $U$ has full dimension and $U \subseteq \bigcap\limits_{t\in N(t_0)} \Pi(t)$. Since $t_0\in T$ was arbitrary, the result follows. 

For (ii), since $\Pi$ is lower hemicontinuous and $\mbox{rint } \Pi(t) \not= \emptyset$ for each $t\in T$, $\mbox{rint } \Pi$ is lower hemicontinuous. 
Also since $\Pi(t)$ is convex, $\mbox{rint }\Pi(t)$ is convex for each $t\in T$. Now let $t_0\in T$ and $\pi_0\in \mbox{rint } \Pi(t_0)$. For each $s\in S$, let $\delta_s\in \Delta(S)$ assign probability 1 to state $s$. Then for each $s=1,\ldots ,S$ there exists $\varepsilon_s>0$ such that $(1-\varepsilon_s)\pi_0 + \varepsilon_s \delta_s \in \mbox{rint }\Pi(t_0)$. Letting $\varepsilon = \min_s \varepsilon_s$, then $\varepsilon >0$ and $(1-\varepsilon) \pi_0 + \varepsilon \delta_s \in \mbox{rint }\Pi(t_0)$ for each $s$.  Then 
\[
\mbox{co } \{ (1-\varepsilon)\pi_0 + \varepsilon \delta_1, \ldots , (1-\varepsilon)\pi_0 + \varepsilon \delta_S \} = \{ (1-\varepsilon)\pi_0 + \varepsilon \pi: \pi \in \Delta(S)\}  \subseteq \mbox{rint }\Pi(t_0)
\]
For each $s$, $(1-\frac{\varepsilon}{2})\pi_0 + \frac{\varepsilon}{2} \delta_s \in \mbox{rint } \mbox{co } \{ (1-\varepsilon)\pi_0 + \varepsilon \delta_1, \ldots , (1-\varepsilon)\pi_0 + \varepsilon \delta_S \} \subseteq \mbox{rint } \Pi(t_0)$. 
 So for each $s$ there exists $\gamma_s>0$ such that $(1-\frac{\varepsilon}{2})\pi_0 + \frac{\varepsilon}{2} \delta_s \in \mbox{rint } \mbox{co } \{ \pi_1, \ldots ,\pi_S \}$ for any $\{\pi_1,\ldots ,\pi_S\}$ such that $\pi_r\in B_{\gamma_s}((1-\varepsilon)\pi_0 + \varepsilon \delta_r)$ for each $r$. Let $\gamma = \min_s \gamma_s >0$. Since $\mbox{rint }\Pi$ is lower hemicontinuous, there exists a neighborhood $N(t_0)$ of $t_0$ such that for every $t\in N(t_0)$ and for each $r=1,\ldots , S$, $\mbox{rint }\Pi(t) \cap B_{\gamma}((1-\varepsilon)\pi_0 + \varepsilon \delta_r)\not= \emptyset$. Since $\mbox{rint } \Pi(t)$ is convex for each $t$, this implies for each $t\in N(t_0)$,  $(1-\frac{\varepsilon}{2})\pi_0 + \frac{\varepsilon}{2} \delta_s \in \mbox{rint } \Pi(t)$ for every $s=1,\ldots ,S$. Using the convexity of $\mbox{rint }\Pi(t)$ again, this implies that for each $t\in N(t_0)$, 
\begin{multline*}
\mbox{co } \left\{ \left(1-\frac{\varepsilon}{2}\right)\pi_0 + \frac{\varepsilon}{2} \delta_1, \ldots , \left(1-\frac{\varepsilon}{2}\right)\pi_0 + \frac{\varepsilon}{2} \delta_S \right\} \\
 =  \left\{ \left(1-\frac{\varepsilon}{2}\right)\pi_0 + \frac{\varepsilon}{2} \pi: \pi \in \Delta(S)\right\} \subseteq \mbox{rint }\Pi(t)
\end{multline*}
Then the set $\{ (1-\frac{\varepsilon}{2})\pi_0 + \frac{\varepsilon}{2} \pi: \pi \in \Delta(S)\}$ has full dimension, and by the argument above $\{ (1-\frac{\varepsilon}{2})\pi_0 + \frac{\varepsilon}{2} \pi: \pi \in \Delta(S)\} \subseteq \bigcap\limits_{t\in N(t_0)} \Pi(t)$. Since $t_0\in T$ was arbitrary, the result follows. 
\end{proof}


Another simple sufficient condition for fully overlapping beliefs builds on the $\varepsilon $-contamination model of Example 1. In this case, for each type $t$ the belief set is $\Pi_\varepsilon (t) = \{ (1-\varepsilon) \pi(t) + \varepsilon \pi: \pi\in \Delta(S)\}$ for a given reference measure $\pi(t) \in \Delta(S)$ and $\varepsilon \in (0,1]$; it is also straightforward to allow the parameter $\varepsilon$ to depend on the type $t$. When the reference measures $\pi(t)$ and $\pi(t')$ are mutually absolutely continuous for all $t, t'$, then beliefs $\{\Pi_{\varepsilon} (t):t\in T\} $ are fully overlapping under some additional continuity conditions on the map from types to reference measures, in particular using properties of the Radon-Nikodym derivatives $\frac{d\pi(t')}{d\pi(t)}$. Notice that as $\varepsilon$ goes to zero in this example, this model converges to the standard Bayesian model. In particular, as this example illustrates, $\Pi (t)$ can be arbitrarily close to a singleton for each $t$, while beliefs $\{\Pi (t) \subseteq \Delta(S):t\in T\}$ are fully overlapping. 

\begin{theorem}
Let $\pi:T\to \Delta(S)$, where $\pi(t)$ and $\pi(t')$ are mutually absolutely continuous for all $t,t'\in T$. Suppose for each $t'\in T$, $\frac{d\pi(t')}{d\pi(t)}$ is continuous in $t$, and the family $\{ \frac{d\pi(t')}{d\pi(t)} : t\in T\}$ is equicontinuous. Then for every $\varepsilon \in (0,1]$, beliefs $\{\Pi_\varepsilon(t) : t\in T\}$ are fully overlapping. 
\end{theorem}
\begin{proof}
Let $\varepsilon \in (0,1]$. Let $t_0\in T$.  By assumption, $\frac{d\pi(t_0)}{d\pi(t)}$ is continuous in $t$, so for each $s\in S$, $\frac{d\pi(t_0)}{d\pi(t)}(s) \to \frac{d\pi(t_0)}{d\pi(t_0)}(s) = 1$  as $t\to t_0$. Then since the family $\{ \frac{d\pi(t_0)}{d\pi(t)} : t\in T\}$ is equicontinuous by assumption, $\frac{d\pi(t_0)}{d\pi(t)} \to 1$ uniformly as $t\to t_0$. 
Then choose a neighborhood $N(t_0)$ of $t_0$ such that for all $t\in N(t_0)$, 
\[
\frac{d\pi(t_0)}{d\pi(t)} (s)  \geq \frac{1-\varepsilon}{1-\frac{\varepsilon}{2}} \ \ \ \forall s\in S
\]
This is possible because $\frac{d\pi(t_0)}{d\pi(t)} \to 1$ uniformly as $t\to t_0$. 

Then note that for every $t\in N(t_0)$, $\Pi_{\frac{\varepsilon}{2}}(t_0) \subseteq \Pi_\varepsilon(t)$. To see this, fix $t\in N(t_0)$. Let $\pi\in \Delta(S)$ and set 
\[
\pi' = \frac{1}{\varepsilon} \left( \frac{\varepsilon}{2} \pi + \left(1-\frac{\varepsilon}{2}\right) \pi(t_0) - (1-\varepsilon)\pi(t)\right)
\]
so that $(1-\frac{\varepsilon}{2})\pi(t_0) + \frac{\varepsilon}{2}\pi = (1-\varepsilon) \pi(t) + \varepsilon \pi'$. Now it suffices to show that $\pi'\in \Delta(S)$. It is straightforward to see that $\pi'$ is additive and $\pi'(S) = 1$, so it suffices to show that $\pi'(A) \geq 0$ for all measurable $A\subseteq S$. To that end, let $A\subseteq S$ be measurable. Then since $t\in N(t_0)$, 
\[
\pi(t_0) (A) = \int_A \frac{d\pi(t_0)}{d\pi(t)} (s) d\pi(t) \geq \left( \frac{1-\varepsilon}{1-\frac{\varepsilon}{2}}\right) \int_A d\pi(t) = \left( \frac{1-\varepsilon}{1-\frac{\varepsilon}{2}} \right) \pi(t) (A)
\]
So
\[
\pi'(A) = \frac{1}{\varepsilon} \left( \frac{\varepsilon}{2} \pi(A) + \left(1-\frac{\varepsilon}{2}\right) \pi(t_0)(A) - (1-\varepsilon)\pi(t)(A)\right) \geq \frac{1}{2} \pi(A) \geq 0
\]
Then $A\subseteq S$ was arbitrary, which implies $\pi'(A) \geq 0$ for all measurable $A\subseteq S$. Thus $\pi'\in \Delta(S)$ and $(1-\frac{\varepsilon}{2})\pi(t_0) + \frac{\varepsilon}{2}\pi = (1-\varepsilon) \pi(t) + \varepsilon \pi' \in \Pi_\varepsilon(t)$. Since $\pi\in \Delta(S)$ was arbitrary, $\Pi_{\frac{\varepsilon}{2}}(t_0) \subseteq \Pi_\varepsilon(t)$. Then $t\in N(t_0)$ was arbitrary, so $\Pi_{\frac{\varepsilon}{2}}(t_0) \subseteq \bigcap\limits_{t\in N(t_0)}\Pi_\varepsilon(t)$, and $\Pi_{\frac{\varepsilon}{2}}(t_0)$ has full dimension. Since $t_0\in T$ was arbitrary, the result follows. 
\end{proof}

\subsection{Quasilinear Model: Main Result}

We are now ready to state and prove the main result of this section. We show that 
robust incentive compatibility can require tight ex post restrictions on the 
payment functions allowed in a robust incentive compatible mechanism. 
In particular,  when beliefs are fully overlapping then robust
incentive compatibility requires the following ex post envelope condition to hold.

\begin{definition}
\label{definition:EPEC quasilinear}A mechanism $(q,p)$ satisfies the \emph{%
ex post envelope condition} if for each $t', t'' \in T$, 
\begin{equation*}
t'' q(t'',s) - p(t'',s) - \left( t' q(t',s) - p(t',s) \right) = \int_{t'}^{t''} q(t,s) dt \ \ \ \forall s\in S
\end{equation*}
\end{definition}

This is a version of ex post revenue equivalence for this screening problem. It requires that the expected payment of a given type be pinned down ex post by the allocation rule $q$ and the ex post payment of the lowest type. That is, equivalently for each type $t'\in T$,
\[
p(t',s) - p(0,s) = t'q(t',s) -\int_{0}^{t'} q(t, s) dt  \ \ \forall s\in S
\]
Notice that this is an ex post rule, rather than the standard interim expected revenue equivalence condition. 

Our main result in this section shows that robustness to sufficiently rich misspecification in beliefs via robust incentive compatibility requires this sharp
form of ex post revenue equivalence. 

\begin{theorem}
\label{theorem:Optimal=>EPEC_ql}If beliefs $\{\Pi (t) \subseteq \Delta(S):t\in T\}$ are fully
overlapping, then any robust incentive compatible mechanism must satisfy
the ex post envelope condition.
\end{theorem}
\begin{proof}
Let $(q,p)$ be a robust incentive compatible mechanism. Fix $t_0 \in T$ and a neighborhood $N(t_0)$ of $t_0$ such that $\bigcap\limits_{t\in N(t_0)} \Pi(t)$ has full dimension. Fix $\pi\in \bigcap\limits_{t\in N(t_0)} \Pi(t)$. For each $t'\in N(t_0)$, set \[
w(t') = \max_{t\in T} E_\pi\left[t'q(t,s) - p(t,s)\right]
\] 
Since $(q,p)$ is robust incentive compatible and $\pi \in \Pi(t')$ for each $t'\in N(t_0)$, $w$ is well defined, and $w(t') = E_\pi[t'q(t',s) - p(t',s)]$. Since $E_\pi[t'q(t,s) - p(t,s)] = t'E_\pi[q(t,s)] - E_\pi[p(t,s)]$ is a continuous convex function of $t'$ for each $t\in T$, $w$ is continuous and convex. Thus $w$ is absolutely continuous and differentiable almost everywhere, and for every $t'$ at which $w$ is differentiable, 
\[
w'(t') = E_\pi[q(t',s)]
\]
Then for any $t', t'' \in N(t_0)$, 
\[
w(t'') - w(t') = \int_{t'}^{t''} E_\pi[q(t,s)] dt
\]
by the fundamental theorem of calculus. 
Then for every $t', t''\in N(t_0)$, 
\[
w(t'') - w(t') = E_\pi[t''q(t'',s) - p(t'',s)] - E_\pi[t'q(t',s) - p(t',s)] = \int_{t'}^{t''} E_\pi[q(t,s)] dt
\]
This implies
\[
E_\pi\left[t''q(t'',s) - p(t'',s) - \left(t'q(t',s) - p(t',s)\right)\right] = E_\pi\left[ \int_{t'}^{t''} q(t,s) dt \right]
\]
using Tonelli's theorem. Since $\pi\in \bigcap\limits_{t\in N(t_0)}\Pi(t)$ was arbitrary, 
\[
E_\pi\left[t''q(t'',s) - p(t'',s) - \left(t'q(t',s) - p(t',s)\right)\right] = E_\pi\left[ \int_{t'}^{t''} q(t,s) dt \right]  \ \ \forall \pi\in \cap_{t\in N(t_0)} \Pi(t)
\]
Since $\bigcap\limits_{t\in N(t_0)} \Pi(t)$ has full dimension, this implies
\[
t''q(t'',s) - p(t'',s) - \left(t'q(t',s) - p(t',s)\right) =  \int_{t'}^{t''} q(t,s) dt  \ \ \ \forall s\in S \tag{$\star$}
\]
Since $t_0\in T$ was arbitrary, this implies for every $t\in T$ there is a neighborhood $N(t)$ of $t$ such that $(\star)$ holds for every $t', t''\in N(t)$. 

Now we claim $(\star)$ holds for every $t', t''\in T$, from the compactness and connectedness of $T$. To see this, note that $\{ N(t): t\in T\}$ is an open cover of $T$ and $T$ is compact, so there exist $t_1, \ldots ,t_n$ such that $T\subseteq \cup_i N(t_i)$. If $T\subseteq N(t_1)$ then we are done, since $(\star)$ holds for all $t', t''\in N(t_1)$. If $T\not\subseteq N(t_1)$, then there must exist $i\in \{ 2,\ldots ,n\}$ such that $N(t_1)\cap N(t_i) \not= \emptyset$; without loss of generality take $t_i = t_2$. Otherwise,  $N(t_1) \cap ( \cup_{i=2}^n N(t_i)) = \emptyset$ and $T\subseteq N(t_1) \cup (\cup_{i=2}^n N(t_i))$, with $T\cap N(t_1) \not= \emptyset$ and $T\cap (\cup_{i=2}^n N(t_i)) \not=\emptyset$,  where $N(t_1)$ and $\cup_{i=2}^n N(t_i)$ are open, which would contradict the connectedness of $T$. Then let $t', t''\in N(t_1) \cup N(t_2)$ and $\bar t \in N(t_1)\cap N(t_2)$. Let $s\in S$. Then
\begin{eqnarray*}
t''q(t'',s) - p(t'',s) - (t'q(t',s) - p(t',s)) &=& t''q(t'',s) - p(t'',s) - (\bar tq(\bar t,s)  - p(\bar t,s)) \\
& \  & \ \ \ \ + \bar tq(\bar t,s) - p(\bar t,s) - (t'q(t',s) - p(t',s))\\
&=& \int_{\bar t}^{t''} q(t,s) dt + \int_{t'}^{\bar t} q(t,s) dt \\
&=& \int_{t'}^{t''} q(t,s) dt
\end{eqnarray*}
Since $s\in S$ was arbitrary, $(\star)$ holds for all $t', t''\in N(t_1) \cup N(t_2)$. 

Then if $T\subseteq N(t_1) \cup N(t_2)$ we are done. If $T\not\subseteq N(t_1) \cup N(t_2)$, then there exists $i\in \{ 3,\ldots ,n\}$ such that $(N(t_1) \cup N(t_2)) \cap N(t_i) \not= \emptyset$; without loss of generality take $t_i = t_3$. Otherwise, $(N(t_1) \cup N(t_2)) \cap (\cup_{i=3}^n N(t_i)) = \emptyset$ and $T\subseteq (N(t_1) \cup N(t_2)) \cup (\cup_{i=3}^n N(t_i))$, with $T\cap \left(N(t_1)\cup N(t_2)\right) \not= \emptyset$ and $T\cap \left(\cup_{i=3}^n N(t_i)\right) \not=\emptyset$, where $N(t_1) \cup N(t_2)$ and $\cup_{i=3}^n N(t_i)$ are open, again contradicting the connectedness of $T$. Then by the above argument, $(\star)$ holds for all $t', t'' \in N(t_1)\cup N(t_2) \cup N(t_3)$. Since $n$ is finite, repeating this argument establishes that $(\star)$ holds for all $t', t''\in T\subseteq \cup_i N(t_i)$, that is, the ex post envelope condition holds. 
\end{proof}

Any robust incentive compatible mechanism must also be robust incentive compatible for the beliefs $\{\overline{\mbox{co}}(\Pi(t)): t\in T\}$, as we noted above. Thus the weaker condition that beliefs $\{\overline{\mbox{co}}(\Pi(t)): t\in T\}$ are fully overlapping is sufficient to imply that any robust incentive compatible mechanism must satisfy the ex post envelope condition. In particular, it is possible that beliefs  $\{\overline{\mbox{co}}(\Pi(t)): t\in T\}$ are fully overlapping yet the sets $\Pi(t)$ and $\Pi(t')$ are disjoint for all $t\not= t'$, so no types share any common beliefs.  Nonetheless, a mechanism that is robust incentive compatible for the underlying beliefs $\{\Pi(t): t\in T\}$ must satisfy the ex post envelope condition.  We record this corollary below. 

\begin{corollary}
Let $\{\Pi(t)\subseteq \Delta(S): t\in T\}$ be given and suppose beliefs $\{\overline{\mbox{co}}(\Pi(t)): t\in T\}$ are fully overlapping. Then any mechanism that is robust incentive compatible for beliefs $\{\Pi(t): t\in T\}$ must satisfy the ex post envelope condition.
\end{corollary}

These results imply that robust incentive compatibility can lead to significant restrictions on the set of feasible
mechanisms. For example, suppose beliefs are fully overlapping and two mechanisms $(q,p)$ and $(q,\tilde{p})
$ are both robust incentive compatible. Then simple manipulations of the ex post 
envelope condition show that there exists a function $k:S\rightarrow {\bf R}$
such that for all $t\in T$, 
\begin{equation*}
p(t,s) - \tilde p(t,s) = k(s)  \ \ \ \forall s\in S
\end{equation*}%
Thus the payment schemes in the two mechanisms can differ at most by a (state-dependent) constant ex post.
A simple consequence of Theorem \ref{theorem:Optimal=>EPEC_ql} is then that
when beliefs are fully overlapping, robust incentive compatible
mechanisms that have the same allocation rule must be revenue equivalent
ex post. We return to draw further implications of this revenue equivalence in auction environments in the next section. 

We show next that for mechanisms using ex post monotone allocation rules, robustness to misspecification of beliefs can select only those that are ex post incentive compatible. To that end, we first define ex post monotonicity.  

\begin{definition}
\label{definition:EPmonotone_allocation}An allocation rule $q$ is \emph{%
ex post monotone} if for all $t', t\in T$,
\begin{equation*}
t' \geq t \Rightarrow q(t',s) \geq q(t,s) \ \ \forall s\in S
\end{equation*}
\end{definition}

Building on Theorem \ref{theorem:Optimal=>EPEC_ql}, we show next that if beliefs are fully overlapping, then
any robust incentive compatible mechanism in which the allocation rule is
ex post monotone must be ex post incentive compatible. 

\begin{theorem}
Suppose beliefs $\{\Pi (t) \subseteq \Delta(S):t\in T\}$ are fully overlapping. Then any robust
incentive compatible mechanism $(q,p)$ in which $q$ is ex post monotone must be ex post
incentive compatible.
\end{theorem}

\begin{proof}
Let $(q,p)$ be a robust incentive compatible mechanism. Since beliefs $\{\Pi (t) \subseteq \Delta(S) :t\in T\}$ are fully overlapping, Theorem \ref%
{theorem:Optimal=>EPEC_ql} implies that $(q,p)$ satisfies the ex post envelope condition. Suppose in addition that $q$ is ex post monotone. Then we will show that the mechanism is also ex post incentive compatible. This is similar to standard arguments for the quasilinear environment; we include the proof for completeness. 

To that end, fix $t''\in T$ and $s\in S$. First consider $t'\in T$ with $t''\geq t'$. Then 
\[
t''q(t'',s) - p(t'',s) - \left(t'q(t',s) - p(t',s)\right) =  \int_{t'}^{t''} q(t,s) dt  \geq \int_{t'}^{t''} q(t',s) dt
\]
where the first equality follows from the ex post envelope condition, and the inequality follows from ex post monotonicity of $q$. Then
\[
t''q(t'',s) - p(t'',s) - \left(t'q(t',s) - p(t',s)\right) \geq \int_{t'}^{t''} q(t',s) dt = q(t',s) \int_{t'}^{t''} dt = (t''-t') q(t',s)
\]
Rearranging, this implies
\[
t''q(t'',s) - p(t'',s) \geq t''q(t',s) -  p(t',s)
\]
Now suppose $t'\geq t''$. Then again using the ex post envelope condition and ex post monotonicity, 
\begin{eqnarray*}
t''q(t'',s) - p(t'',s) - \left(t'q(t',s) - p(t',s)\right) =  \int_{t'}^{t''} q(t,s) dt 
&=& - \int_{t''}^{t'} q(t,s) dt\\
&\geq& - \int_{t''}^{t'} q(t',s) dt = (t'' - t') q(t', s)
\end{eqnarray*}
Rearranging again yields
\[
t''q(t'',s) - p(t'',s) \geq t''q(t',s) -  p(t',s)
\]
Since $t''\in T$ and $s\in S$ were arbitrary, for all $t''\in T$, 
\[
t''q(t'',s) - p(t'',s) \geq t''q(t',s) -  p(t',s) \ \ \ \forall t'\in T \mbox{ and } \forall s\in S
\]
Thus $(q,p)$ is ex post incentive compatible. 
\end{proof}

Notice that for a given ex post monotone allocation rule $q$, the importance
of this result is \emph{not} that it is possible to implement this
allocation rule in an ex post incentive compatible way. This is well-known,
and is simply done with a payment rule satisfying the ex post envelope
condition. Instead, this result shows that whenever beliefs are fully
overlapping, these are the \emph{only} payment schemes that can implement
this allocation rule in a robust incentive compatible way. In contrast, in a standard Bayesian setting a seller might have significantly more flexibility while
satisfying interim incentive compatibility and ex post monotonicity. In this sense, requiring mechanisms to be robust to misspecification of beliefs can generate a discontinuity in the set of feasible mechanisms and uniquely select those that are ex post incentive compatible. 

To develop a deeper understanding of the role of  monotonicity in these results, we close the section by examining the extent to which some form of monotonicity is necessary for robust incentive compatibility. We start with a simple observation.

\begin{lemma}
\label{lemma:_monotonicity_of_allocation_rule} If a mechanism $(q,p)$
is robust incentive compatible for beliefs $\{\Pi(t)\subseteq \Delta(S): t\in T\}$, then for all $t,t'\in T$ with $t'\geq t$, 
\[
E_{\pi
}[q(t',s)]\geq E_{\pi }[q(t,s)]   \ \ \ \forall \pi \in \Pi (t^{\prime
})\cap \Pi (t)
\]
\end{lemma}

\begin{proof}
By robust incentive compatibility, for any $t,t^{\prime }\in T$,%
\[
E_{\pi }[tq(t,s)-p(t,s)]  \geq E_{\pi }\left[ tq(t^{\prime },s)-p(t^{\prime
},s)\right] \ \ \forall \pi \in \Pi (t) 
\]
and 
\[
E_{\pi }[t^{\prime }q(t^{\prime },s)-p(t^{\prime },s)] \geq E_{\pi }\left[
t^{\prime }q(t,s)-p(t,s)\right] \ \ \forall \pi \in \Pi (t^{\prime })
\]
If $\pi \in \Pi (t^{\prime })\cap \Pi (t)$, this implies
\begin{equation*}
E_{\pi }[t'q(t',s)-tq(t',s)]\geq E_{\pi }\left[ t'q(t,s)-tq(t,s)\right] 
\end{equation*}
or 
\begin{equation*}
(t'-t)E_{\pi }[q(t',s)]\geq (t'-t)E_{\pi }\left[
q(t,s)\right] 
\end{equation*}
If $t'\geq t$, this implies $E_\pi [q(t',s)] \geq E_\pi[q(t,s)%
] $. Since $\pi\in \Pi (t^{\prime })\cap \Pi (t)$ was arbitrary, the result follows.
\end{proof}

This result shows that whenever two types share a common belief, then robust incentive compatibility requires that the expected value of the allocation
rule taken with respect to this common belief must be larger for the higher type. In a standard Bayesian model, this observation is at the heart of the fact that interim monotonicity is necessary for interim incentive compatibility when beliefs are independent of types. In this case, beliefs are single-valued for each type, so for each type $t$, $\Pi(t) = \{ \pi(t)\}$ for some $\pi(t)\in \Delta(S)$. In addition, beliefs are independent of types, so $\pi(t) = \pi(t') = \pi \in \Delta(S)$ for any pair of types $t,t'$. In contrast, for a Bayesian model with beliefs that can depend on types, interim monotonicity is no longer necessary for interim incentive compatibility. Thus the standard Bayesian model collapses conditions that are distinct in our model, including that types share common beliefs and that beliefs are independent of types. In our model many different conditions reflect similar restrictions, and each has different implications regarding the sense in which monotonicity is necessary for incentive compatibility. To better understand the roles these different conditions play, we explore a variety of them here.

For the remaining discussion, consider a fixed mechanism $(q,p)$ that is robust incentive compatible for beliefs $\{ \Pi(t) \subseteq \Delta(S): t\in T\}$. 
We say types share a {\it common belief} if $\bigcap\limits_{t\in T}\Pi (t)\not=\emptyset $. In this case, for any such common belief $\pi \in \bigcap\limits_{t\in T}\Pi (t)$, $E_\pi[q(\cdot , s)]$ must be increasing in $t$, by Lemma 2. We say beliefs are {\it independent} if $\Pi(t) = \Pi \subseteq \Delta(S)$ for all $t \in T$. Independence is clearly stronger than common belief, and has the stronger implication that for any belief of any type $\pi\in \Pi$,  $E_\pi[q(\cdot , s)]$ must be increasing in $t$. 
 This restriction can be thought of as a robust and
global version of interim monotonicity: interim monotonicity must hold for
all beliefs of all types. 

Similarly, local versions of
independence impose local monotonicity restrictions. We say {\it beliefs are overlapping} if 
for every type $t'$ there is a neighborhood $%
N(t')$ of $t'$ such that $\bigcap\limits_{t\in N(t')}\Pi (t)\not=\emptyset $. This expresses a local version of independence. When beliefs are overlapping, robust incentive compatibility requires that the allocation rule
must satisfy a local version of robust monotonicity. We say that $q$ is {\it locally robustly monotone} if for all types $t'$ there exists a nonempty set $\Pi\subseteq \Delta (S)$ and a
neighborhood $N(t')$ of $t'$ such that $E_{\pi }[q(\cdot,s)]$ is an increasing
function of $t$ on $N(t')$ for all $\pi \in \Pi$. Using this terminology, if beliefs are overlapping, $q$ must be locally robustly monotone. 

Notice that as the set of common beliefs increases (in the sense of set inclusion), the monotonicity restrictions imposed on the allocation rule $q$ become sharper. In the limiting case, in which $\Pi (t)=\Delta (S)$ for all $t$,
 $q$ must be ex post monotone.  Conversely, for any 
allocation rule $q$ that is not ex post monotone,  there exists $\Pi \not=\Delta
(S)$ sufficiently large such that if $\Pi \subseteq \bigcap\limits_{t\in T}\Pi (t) $
then $q$ is not part of any robust incentive compatible mechanism for these beliefs. 

We collect all of these results below. We omit the straightforward proofs. 

\begin{theorem}
\label{theorem:_monotonicity_of_allocation_rule}
Let $(q,p)$ be a robust incentive compatible mechanism for beliefs $\{ \Pi(t) \subseteq \Delta(S): t\in T\}$. 
\begin{itemize}
  \item[(i)] If types share a common belief, then  $E_\pi[q(\cdot , s)]$ is increasing in $t$ for all $\pi \in \bigcap\limits_{t\in T}\Pi (t)$.
  \item[(ii)] If beliefs are independent, with $\Pi(t) = \Pi\subseteq \Delta(S)$ for all $t\in T$, then $E_\pi[q(\cdot , s)]$ is increasing in $t$ for all $\pi \in  \Pi$.
  \item[(iii)] If beliefs are overlapping, then $q$ is locally robustly monotone. 
  \item[(iv)] If $\Pi(t) = \Delta(S)$ for all $t \in T$, then $q$ is ex post monotone.
 \end{itemize}
Finally, if $q$ is not ex post monotone, then  there exists $\Pi \not=\Delta
(S)$ sufficiently large such that $(q,p)$ is not robust incentive compatible for any payment function $p$ and any beliefs $\{ \Pi(t) \subseteq \Delta(S): t\in T\}$ with $\Pi \subset \bigcap\limits_{t\in T}\Pi (t)$. 
\end{theorem}

We conclude this section with an example illustrating the necessary role of fully overlapping
beliefs in the equivalence between robust incentive compatibility and ex post incentive compatibility. The example gives a robust incentive compatible mechanism that is not ex post incentive compatible, and in which the ex post envelope condition fails. 

\noindent {\bf Example 2: } 
Let $S=\{1,2\}$.  Let $\varepsilon \in (0,\frac{1}{6}]$, and let
\begin{equation*}
\Pi(t)=
\begin{cases}
\{ (\pi_1,\pi_2)\in \Delta(S): \pi_1 \in [ \frac{1}{3}-2\varepsilon ,\frac{1}{3}-\varepsilon ] \}& \mbox{ if } 0\leq t\leq \frac{1}{2} \\ 
\{ (\pi_1,\pi_2) \in \Delta(S): \pi_1 \in [ \frac{2}{3}+\varepsilon ,\frac{2}{3}+2\varepsilon ] \} & \mbox{ if } \frac{1%
}{2}<t\leq 1%
\end{cases}%
\end{equation*}%
Let $c>0$. Consider the mechanism $(q,p)$, where $q(t ,s)=1$ for all $(t ,s)$ and 
\begin{equation*}
p(t ,s)=
\begin{cases}
\frac{1}{2}c &   \text{ if } t\leq \frac{1}{2} \text{ and } s=1 \\ 
-c &  \text{ if } t \leq \frac{1}{2}\text{ and } s=2 \\ 
0 &   \text{ if } t > \frac{1}{2}%
\end{cases}%
\end{equation*}%
Then  the interim expected utility function for type $t$, for a given $\pi\in \Pi(t)$, is
\begin{equation*}
E_{\pi} \left[ tq(\theta ,s)-p(\theta ,s) \right]=
\begin{cases}
t+c(1-\frac{3}{2}\pi_1 )   & \text{ if } \theta \leq \frac{1}{2} \\ 
t &   \text{ if } \theta > \frac{1}{2}%
\end{cases}
\end{equation*}%
This is maximized by any report $\theta \leq \frac{1}{2}$ if $\pi_1 <\frac{2}{3}$, and by any report $%
\theta > \frac{1}{2}$ if $\pi_1 >\frac{2}{3}$. Since $\pi_1 <\frac{2}{3}$ for
all $\pi \in \Pi (t)$ when $t\leq \frac{1}{2}$, and $\pi_1 >\frac{2}{3}$ for
all $\pi \in \Pi (t)$ when $t>\frac{1}{2}$, the mechanism is  robust incentive compatible. In addition, the allocation rule $q$ is clearly ex post monotone, as it is constant. This mechanism is not
ex post incentive compatible, however. Since $c>0$, when $s=2$ any type $t>\frac{1}{2}$ prefers
to report $\theta \leq \frac{1}{2}$. Here $\Pi(t)$ has full dimension for each type $t\in T$, but beliefs are not fully overlapping, as for $t=\frac{1}{2}$, $\Pi(t) \cap \Pi(t') = \emptyset$ for all $t'>t$. \hfill$\diamondsuit$

\section{\label{section:Auctions}Robust Incentive Compatible Auctions and Revenue Equivalence}

In this section, we illustrate how to extend the main results of the previous section from a single agent setting to a multi-agent setting by considering
a simple common value auction. We extend the main result of the previous section by showing that when beliefs are fully overlapping, auctions that satisfy robust incentive compatibility must also satisfy an ex post envelope condition. As a
consequence, in the large class of ex post monotone auction mechanisms, only the
ex post incentive compatible mechanisms are robustly incentive compatible in this case. For example, 
although both first-price and second-price auctions are ex post monotone, only 
second-price auctions are ex post incentive compatible, and thus robustly incentive compatible when beliefs are fully overlapping. We also show that a robust revenue equivalence 
result holds in this case: the ex post payment functions of two mechanisms that satisfy robust 
incentive compatibility and allocate the object in the same way can differ at most by
a constant.

The setup is similar to the previous section, but now admits many buyers by setting
the state space of each buyer to be the set of type profiles of the other buyers. The owner of an indivisible object considers
selling it to one of many potential buyers using an auction mechanism.  The set
of potential buyers is $I=\left\{ 1,\ldots ,n\right\} $. Buyers are privately informed about their valuations for the object. The type of buyer $i$, denoted $%
t_{i}$, is drawn from $T_{i}=[0,1]$, and his valuation for the good is given
by the function $v_{i}:T_{i}\times T_{-i}\rightarrow \mathbf{R}$.\footnote{%
Following usual conventions, the subscript $-i$ is used for variables that
pertain to all players except $i$.} Thus values can be interdependent. 

We make use of the following conditions on valuation functions throughout this section. 

\begin{assumption}
\label{Assumption:auctions_v_differentiable} For each $i\in I$, $v_{i}:T_{i}\times T_{-i}\rightarrow \mathbf{R}$ is
non-decreasing and continuously differentiable with respect to $t_i$, with $\frac{\partial v_{i}}{%
\partial t_{i}} (t_{i},t_{-i}) \geq 0$ for all $t_{i}\in T_{i},$ $%
t_{-i}\in T_{-i}$.
\end{assumption}

We let  $q_{i} \in [0,1]$ denote the probability that buyer $i$ receives the good, and $%
p_{i} \in {\bf R}$ denote the payment he makes to the seller. Given $(q_i,p_i)$ and a realized profile of types $(t_i, t_{-i})$, buyer $i$'s utility is 
then $v_{i}(t_{i},t_{-i})q_{i}-p_{i}$. 

A direct mechanism consists of $2n$ functions $( q,p)
=(q_{i},p_{i})_{i\in I}$, with $q:\prod_i T_i \to \Delta (I\cup \{0\})$ where 0 denotes the outcome in which the object remains with the seller. Here $(q_{i},p_{i})$ are buyer $i$'s allocation
rule and payment function as a function of the profile of buyers' reports. So   
$q_{i}:T_{i}\times T_{-i}\rightarrow [0,1]$ gives the probability that buyer $i$ receives the object, and $p_{i}:T_{i}\times
T_{-i}\rightarrow \mathbf{R}$ specifies the payment that buyer $i$ makes to
the seller. We assume that $q_i(t_i, \cdot)$ and $p_i(t_i, \cdot)$ are Borel measurable for each $t_i\in T_i$, and each $i\in I$. 
For any profiles of true types $(t_{i},t_{-i})\in T_{i}\times T_{-i}$ and
reported types $(\theta _{i},\theta _{-i})\in T_{i}\times T_{-i}$, the
ex post utility of buyer $i$ is then
\begin{equation*}
v_{i}(t_{i},t_{-i})q_{i}(\theta _{i},\theta _{-i})-p_{i}(\theta _{i},\theta
_{-i})
\end{equation*}%
This is a standard interdependent values auction model as presented,
for example, in \cite{Krishna02}.

Next we adapt definitions of robust, interim, and ex post incentive compatibility from the
previous section to allow for multiple agents. Since there are many agents, each incentive compatibility condition now reflects equilibrium behavior. 

As in the previous section, we start with the most restrictive notion. Ex post incentive compatibility requires  incentive compatibility constraints to hold for any possible realization of the state. Since the state space here is the type space of the other players, this requires truthful reporting to be a dominant strategy equilibrium in the direct mechanism. 

\begin{definition}
A mechanism $(q,p)$ is \emph{ex post incentive compatible} if for each
buyer $i\in I$ and for each $t_{i},\theta _{i}\in T_{i}$,%
\[
v_{i}(t_{i},t_{-i})q_{i}( t_{i},t_{-i}) -p_{i}(
t_{i},t_{-i}) \geq v_{i}(t_{i},t_{-i})q_{i}( \theta _{i},t_{-i})
-p_{i}( \theta _{i},t_{-i}) \ \  \ \forall
t_{-i}\in T_{-i}
\]
\end{definition}

Interim incentive compatibility here is again standard, and requires truthful reporting to be a Bayes-Nash equilibrium for the corresponding belief profile.

\begin{definition}
A mechanism $(q,p)$ is \emph{interim incentive
compatible} given beliefs $\{\pi (t_{i}) \in \Delta(T_{-i}) :t_{i}\in T_{i}\}$ for each $i\in I$ if for each buyer $i\in I$ and for each $t_{i}, \theta_i\in
T_{i}$,
\[
E_{\pi(t_i) }\left[ v_{i}(t_{i},t_{-i})q_{i}( t_{i},t_{-i})
-p_{i}( t_{i},t_{-i}) \right] \geq E_{\pi(t_i) }%
\left[ v_{i}(t_{i},t_{-i})q_i(\theta _{i},t_{-i})-p_i(\theta
_{i},t_{-i})\right] 
\]
\end{definition}

Finally, robust incentive compatibility requires interim  incentive compatibility to be robust to
misspecification of beliefs.

\begin{definition}
A mechanism $(q,p)$ is \emph{robust incentive compatible} for beliefs $\{\Pi
(t_{i}) \subseteq \Delta(T_{-i}):t_{i}\in T_{i}\}$ for each $ i\in I$ if for each buyer $i\in I$ and for each $t_{i}, \theta_i\in
T_{i}$, 
\[
E_{\pi }\left[ v_{i}(t_{i},t_{-i})q_{i}( t_{i},t_{-i})
-p_{i}( t_{i},t_{-i}) \right] \geq E_{\pi }%
\left[  v_{i}(t_{i},t_{-i})q_i(\theta _{i},t_{-i})-p_i(\theta
_{i},t_{-i})\right] \ \ \forall \pi \in \Pi (t_{i})
\label{eq:auctionsOptimalIC}
\]
\end{definition}

The relationship between these notions is analogous to the single agent case. 
In particular, the three notions are nested: an ex post incentive compatible mechanism is also 
interim and robust incentive compatible, and a robust incentive compatible mechanism is interim 
incentive compatible (provided $\pi(t_{i})\in \Pi (t_{i})$ for each $t_{i}\in T_{i}$ and for each $i$). When beliefs are not
misspecified, each element of $\{\Pi (t_{i}) \subseteq \Delta(T_{-i}):t_{i}\in T_{i}\}$ is a singleton and robust incentive compatibility
reduces to interim incentive compatibility. At the other extreme, if beliefs are completely 
misspecified and $\Pi
(t_{i})=\Delta(T_{-i})$ for each $t_{i}\in T_{i}$ and each $i$, then robust incentive
compatibility is equivalent to ex post incentive compatibility, by an application of Lemma 1.

The first result of this section echoes the main result of the previous section. When beliefs are fully overlapping, robust incentive compatible mechanisms must satisfy an ex post envelope condition. To formalize this result, we begin by adapting the ex post envelope condition from the previous section to this multi-agent setting.

\begin{definition}
\label{definition:EPEC quasilinear auction}A mechanism $(q,p)$ satisfies the 
\emph{ex post envelope condition} if for each $i\in I$ and for each $%
t_i', t_i''\in T_i$, 
\begin{multline*}
v(t_i'', t_{-i})q(t_i'', t_{-i}) - p(t_i'', t_{-i}) - \left(v(t_i', t_{-i})q(t_i', t_{-i}) - p(t_i', t_{-i}) \right) \\ =  \int_{t_i'}^{t_i''} \frac{\partial v_i}{\partial t_i}(t_i, t_{-i})q_i(t_i,t_{-i}) dt_i  \ \ \ \forall t_{-i} \in T_{-i}
\end{multline*}
\end{definition}

We give the main result of this section next. The argument mimics the proof of Theorem \ref{theorem:Optimal=>EPEC_ql},
modified to take into account that each agent's valuation function depends on the other agents' types in this multi-agent setting. 

\begin{theorem}
\label{theorem:OIC=>EPEC_auctions}If Assumption \ref%
{Assumption:auctions_v_differentiable} holds and beliefs $\{\Pi (t_{i}) \subseteq \Delta(T_{-i}):t_{i}\in
T_{i}\}$ are fully overlapping for each $i\in I$, then any robust incentive
compatible mechanism satisfies the ex post envelope condition.
\end{theorem}

\begin{proof}
Let $(q,p)$ be a robust incentive compatible mechanism. Fix $i\in I$. Fix $t_0\in T_i$ and a neighborhood $N(t_0)$ of $t_0$ such that $\bigcap\limits_{t_i\in N(t_0)} \Pi(t_i)$ has full dimension. Fix $\pi\in \bigcap\limits_{t_i\in N(t_0)} \Pi(t_i)$. For each $t_i'\in N(t_0)$, set 
\[
w_i(t_i') = \max_{t_i\in T_i} E_\pi \left[ v(t_i', t_{-i})q(t_i, t_{-i}) - p(t_i, t_{-i})\right]
\]
Since $(q,p)$ is robust incentive compatible and $\pi \in \Pi(t_i')$ for each $t_i'\in N(t_0)$, $w_i$ is well defined, and $w_i(t_i') = E_\pi\left[ v(t_i', t_{-i})q(t_i', t_{-i}) - p(t_i', t_{-i})\right]$. By Assumption 1, $v_i$ is uniformly Lipschitz continuous in $t_i$, and $q_i(t_i, t_{-i}) \in [0,1]$ for all $(t_i, t_{-i})$. Thus for each $t_i\in T_i$, $ E_\pi \left[ v(t_i', t_{-i})q(t_i, t_{-i}) - p(t_i, t_{-i})\right]$ is uniformly Lipschitz continuous in $t_i'$. This implies $w_i$ is Lipschitz continuous, and thus absolutely continuous and differentiable almost everywhere. By a version of the envelope theorem for Lipschitz functions due to Clarke (\cite{Clarke}, Theorem 2.8.6), for every $t_i'$ at which $w_i$ is differentiable, 
\[
w_i'(t_i') = E_\pi\left[ \frac{\partial v_i}{\partial t_i}(t_i', t_{-i})q_i(t_i',t_{-i})\right]
\]
Then for any $t_i', t_i''\in N(t_0)$, 
\[
w_i(t_i'') - w_i(t_i') = \int_{t_i'}^{t_i''} E_\pi\left[ \frac{\partial v_i}{\partial t_i}(t_i, t_{-i})q_i(t_i,t_{-i})\right] dt_i
\]
by the fundamental theorem of calculus. 
Thus 
\begin{eqnarray*}
w_i(t_i'') - w_i(t_i') &=& E_\pi \left[v(t_i'', t_{-i})q(t_i'', t_{-i}) - p(t_i'', t_{-i})\right] - E_\pi\left[v(t_i', t_{-i})q(t_i', t_{-i}) - p(t_i', t_{-i})\right] \\
&=& \int_{t_i'}^{t_i''} E_\pi\left[ \frac{\partial v_i}{\partial t_i}(t_i, t_{-i})q_i(t_i,t_{-i})\right] dt_i
\end{eqnarray*}
Using Tonelli's theorem, this implies
\begin{multline*}
E_\pi \left[v(t_i'', t_{-i})q(t_i'', t_{-i}) - p(t_i'', t_{-i}) - \left(v(t_i', t_{-i})q(t_i', t_{-i}) - p(t_i', t_{-i}) \right) \right] \\ = E_\pi\left[ \int_{t_i'}^{t_i''} \frac{\partial v_i}{\partial t_i}(t_i, t_{-i})q_i(t_i,t_{-i}) dt_i \right]
\end{multline*}
Since $\pi \in \bigcap\limits_{t_i\in N(t_0)} \Pi(t_i)$ was arbitrary, 
\begin{multline*}
E_\pi\left[v(t_i'', t_{-i})q(t_i'', t_{-i}) - p(t_i'', t_{-i}) - \left(v(t_i', t_{-i})q(t_i', t_{-i}) - p(t_i', t_{-i}) \right) \right] \\  = E_\pi\left[ \int_{t_i'}^{t_i''} \frac{\partial v_i}{\partial t_i}(t_i, t_{-i})q_i(t_i,t_{-i}) dt_i \right] \ \ \forall \pi\in \bigcap\limits_{t_i\in N(t_0)} \Pi(t_i)
\end{multline*}
Then since $\bigcap\limits_{t_i\in N(t_0)} \Pi(t_i)$ has full dimension, 
\begin{multline}
v(t_i'', t_{-i})q(t_i'', t_{-i}) - p(t_i'', t_{-i}) - (v(t_i', t_{-i})q(t_i', t_{-i}) - p(t_i', t_{-i})) \\  =  \int_{t_i'}^{t_i''} \frac{\partial v_i}{\partial t_i}(t_i, t_{-i})q_i(t_i,t_{-i}) dt_i \ \ \forall t_{-i} \in T_{-i}  \tag{$\star\star$}
\end{multline}
Since $t_i', t_i'' \in N(t_0)$ were arbitrary, $(\star\star)$ holds for every $t_i', t_i''\in N(t_0)$. Since $t_0\in T_i$ was arbitrary, for every $t_i\in T_i$ there exists a neighborhood $N(t_i)$ of $t_i$ such that $(\star\star)$ holds for all $t_i', t_i'' \in N(t_i)$. Now from the compactness and connectedness of $T_i$, as in the proof of Theorem 3, $(\star\star)$ holds for all $t_i', t_i'' \in T_i$. Repeating this argument for each $i\in I$ establishes the result. 
\end{proof}

As in the single agent case, this result suggests that robust incentive compatibility can impose stringent restrictions on mechanisms. 
 For example, consider an auction in which the allocation rules are ex post monotone, that is, such that $q_i$ is non-decreasing in $t_i$ for each buyer $i$. When beliefs are fully overlapping and our basic assumptions hold, then such an auction can be robust incentive compatible only if it is ex post incentive compatible.

\begin{theorem}
\label{theorem:OIC+EPM=>EPIC_auctions}Suppose Assumption 1 holds and beliefs $\{\Pi (t_{i}) \subseteq \Delta(T_{-i}):t_{i}\in
T_{i}\} $ are fully overlapping for each $i\in I$. If $(q,p)$ is a robust
incentive compatible mechanism and $q_{i}$ is non-decreasing in $t_i$ for each $i$, then $(q,p)$ is ex post incentive compatible.
\end{theorem}

\begin{proof}
Fix $i\in I$. By Theorem \ref{theorem:OIC=>EPEC_auctions}, $(q,p)$ must
satisfy the ex post envelope condition. We claim that when $q$ is ex post monotone, 
$(q,p)$ must be ex post incentive compatible for agent $i$. As with Theorem 4, the argument follows standard results for the quasilinear setting; we include the proof for completeness. 

To that end, fix $t_i'' \in T_i$ and $t_{-i} \in T_{-i}$. First consider $t_i'\in T_i$ with $t_i'' \geq t_i'$. Then 
\begin{eqnarray*}
v(t_i'', t_{-i})q(t_i'', t_{-i}) - p(t_i'', t_{-i}) -\left(v(t_i', t_{-i})q(t_i', t_{-i}) - p(t_i', t_{-i}) \right) 
& =&  \int_{t_i'}^{t_i''} \frac{\partial v_i}{\partial t_i}(t_i, t_{-i})q_i(t_i,t_{-i}) dt_i \\
& \geq & \int_{t_i'}^{t_i''} \frac{\partial v_i}{\partial t_i}(t_i, t_{-i})q_i(t_i',t_{-i}) dt_i 
\end{eqnarray*}
where the first equality follows from the ex post envelope condition, and the inequality follows from the ex post monotonicity of $q_i$. By the fundamental theorem of calculus, 
\begin{eqnarray*}
v(t_i'', t_{-i})q(t_i'', t_{-i}) - p(t_i'', t_{-i}) - \left(v(t_i', t_{-i})q(t_i', t_{-i}) - p(t_i', t_{-i}) \right) 
&  \geq & \int_{t_i'}^{t_i''} \frac{\partial v_i}{\partial t_i}(t_i, t_{-i})q_i(t_i',t_{-i}) dt_i \\
& = & \left( v_i(t_i'', t_{-i}) - v_i(t_i', t_{-i}) \right)q_i(t_i',t_{-i}) 
\end{eqnarray*}
Rearranging, this implies 
\[
v(t_i'', t_{-i})q(t_i'', t_{-i}) - p(t_i'', t_{-i})  \geq  v_i(t_i'', t_{-i})q_i(t_i',t_{-i}) - p(t_i', t_{-i})
\]
Now suppose $t_i'\geq t_i''$. Then again using the ex post envelope condition and ex post monotonicity of $q_i$, 
\begin{eqnarray*}
v(t_i'', t_{-i})q(t_i'', t_{-i}) - p(t_i'', t_{-i}) - \left(v(t_i', t_{-i})q(t_i', t_{-i}) - p(t_i', t_{-i}) \right) 
& = &  \int_{t_i'}^{t_i''} \frac{\partial v_i}{\partial t_i}(t_i, t_{-i})q_i(t_i,t_{-i}) dt_i \\
& = & -\int_{t_i''}^{t_i'} \frac{\partial v_i}{\partial t_i}(t_i, t_{-i})q_i(t_i,t_{-i}) dt_i \\
&  \geq &- \int_{t_i''}^{t_i'} \frac{\partial v_i}{\partial t_i}(t_i, t_{-i})q_i(t_i',t_{-i}) dt_i 
\end{eqnarray*}
By the fundamental theorem of calculus, 
\begin{eqnarray*}
v(t_i'', t_{-i})q(t_i'', t_{-i}) - p(t_i'', t_{-i}) - \left(v(t_i', t_{-i})q(t_i', t_{-i}) - p(t_i', t_{-i}) \right)
 &\geq& -\int_{t_i''}^{t_i'} \frac{\partial v_i}{\partial t_i}(t_i, t_{-i})q_i(t_i',t_{-i}) dt_i \\
&=& \left( v_i(t_i'', t_{-i}) - v_i(t_i', t_{-i}) \right)q_i(t_i',t_{-i}) 
\end{eqnarray*}
Rearranging again, this implies 
\[
v(t_i'', t_{-i})q(t_i'', t_{-i}) - p(t_i'', t_{-i})  \geq  v_i(t_i'', t_{-i})q_i(t_i',t_{-i}) - p(t_i', t_{-i})
\]
Since $t_i''\in T_i$ and $t_{-i} \in T_{-i}$ were arbitrary, for all $t_i''\in T_i$,
\[
v(t_i'', t_{-i})q(t_i'', t_{-i}) - p(t_i'', t_{-i})  \geq  v_i(t_i'', t_{-i})q_i(t_i',t_{-i}) - p(t_i', t_{-i}) \ \ \ \forall t_i'\in T_i \mbox{ and } \forall t_{-i} \in T_{-i}
\]
Thus $(q,p)$ is ex post incentive compatible for agent $i$. Repeating this argument for each $i\in I$ shows that $(q,p)$ is ex post incentive compatible. \end{proof}

The strength of this result is highlighted by observing that although many auction mechanisms 
satisfy ex post monotonicity, including first and second price auctions, few of these also satisfy 
the ex post envelope condition. For example, the first-price auction does not, and thus would be 
ruled out by robust incentive compatibility whenever beliefs are fully overlapping. 

This result also has sharp implications for ex post revenue equivalence. 
Consider two robust incentive compatible mechanisms $(q,p)$ and $(q,\tilde{p%
})$. Note that the allocation rule $q$ is the same so these mechanisms assign the
object in the same way. Suppose beliefs $\{ \Pi(t_i) \subseteq \Delta(T_{-i}): t_i\in T_i\}$ are fully overlapping for each $i \in I$. As a consequence of the ex post envelope condition, the payment functions for a given agent in these mechanisms can  differ only by an amount that is independent of his report. That is, for a given agent $i$,
\[
p_i(t_i, t_{-i}) -\tilde p_i(t_i, t_{-i}) = k_i(t_{-i})
\]
for some function $k_i:T_{-i}\to {\bf R}$. Now suppose in addition that for some type $t_i$, agent $i$ is robustly indifferent between the two mechanisms, in the sense that his expected utility in either mechanism is the same for all of his beliefs. That is, 
\[
E_{\pi }\left[ v_{i}(t_{i},t_{-i})q_{i}( t_{i},t_{-i})
-p_{i}( t_{i},t_{-i}) \right] = E_{\pi }\left[ v_{i}(t_{i},t_{-i})q_i(t_{i},t_{-i})-\tilde p_i(t_{i},t_{-i})\right] \ \ \forall \pi \in \Pi (t_{i})
\]
For example, $t_i$ might be the lowest type for agent $i$, who receives a constant reservation utility in both mechanisms. Then 
\[
E_{\pi }\left[ p_{i}( t_{i},t_{-i}) \right] = E_{\pi }\left[ \tilde p_i(t_{i},t_{-i})\right] \ \ \forall \pi \in \Pi (t_{i})
\]
This implies
\[
E_\pi [k_i(t_{-i})] = 0 \ \ \forall \pi\in \Pi(t_i)
\]
In this case, $k_i \equiv 0$ because $\Pi(t_i)$ has full dimension. Thus the payment functions $p_i$ and $\tilde p_i$ must be identical, and uniquely determined by the allocation rule $q_i$. 

We record these observations below.

\begin{theorem}
\label{theorem:expost_revenue_equivalence} Suppose Assumption 1 holds and beliefs $\{ \Pi(t_i) \subseteq \Delta(T_{-i}): t_i\in T_i\}$ are fully overlapping for each $i \in I$. If $( q,p) $ and $( q,\tilde{p}) $ are robust
incentive compatible mechanisms, then for each $i\in I$ there exists a
function $k_{i}:T_{-i}\rightarrow {\bf R}$ such that for all $t_{i}\in T_{i}$,
\begin{equation*}
p_{i}( t_{i},t_{-i}) -\tilde{p}_{i}( t_{i},t_{-i}) = k_{i} (t_{-i})  \ \ \  \forall t_{-i}\in T_{-i}
\end{equation*}%
Moreover, if for agent $i$ there exists a type $t_{i}^{\ast } \in T_i$ who is robustly indifferent between these mechanisms, that is, such that 
\[
E_{\pi }\left[ v_i(t^*_i, t_{-i}) q_{i}( t^*_{i},t_{-i})
-p_{i}( t^*_{i},t_{-i}) \right] = E_{\pi }\left[ v_{i}(t^*_{i},t_{-i})q_i(t^*_{i},t_{-i})-\tilde p_i(t^*_{i},t_{-i})\right] \ \ \forall \pi \in \Pi (t^*_{i})
\]
then $k_{i}\equiv 0$.
\end{theorem}

\begin{proof}
By Theorem \ref{theorem:OIC=>EPEC_auctions}, $(q,p)$ and $(q,\tilde p)$ must both satisfy the ex post envelope condition. Then fix $i\in I$ and $t_i' \in T_i$. By the ex post envelope condition for $(q,p)$, 
\begin{multline*}
p_{i}( t_{i}',t_{-i}) -p_{i}( 0,t_{-i}) =v_{i}(t_{i}',t_{-i})q_{i}(
t_{i}',t_{-i}) -v_{i}(0,t_{-i})q_{i}( 0,t_{-i}) \\
-\int_{0}^{t_{i}'}\frac{\partial v_{i}}{\partial t_{i}}(t_i, t_{-i})q_{i}( t_i ,t_{-i}) dt_i  \ \ \ \forall t_{-i} \in T_{-i}
\end{multline*}%
Similarly, the ex post envelope condition for $(q,\tilde p)$ implies 
\begin{multline*}
\tilde p_{i}( t_{i}',t_{-i}) - \tilde p_{i}( 0,t_{-i}) =v_{i}(t_{i}',t_{-i})q_{i}(
t_{i}',t_{-i}) -v_{i}(0,t_{-i})q_{i}( 0,t_{-i}) \\
-\int_{0}^{t_{i}'}\frac{\partial v_{i}}{\partial t_{i}}(t_i, t_{-i})q_{i}( t_i ,t_{-i}) dt_i  \ \ \ \forall t_{-i} \in T_{-i}
\end{multline*}
Thus
\[
p_i(t_i', t_{-i}) - p_i(0, t_{-i}) = \tilde p_i(t_i', t_{-i}) - \tilde p_i(0, t_{-i}) \ \ \ \forall t_{-i} \in T_{-i}
\]
or equivalently,
\[
p_i(t_i', t_{-i}) -  \tilde p_i(t_i', t_{-i})=  p_i(0, t_{-i})- \tilde p_i(0, t_{-i}) \ \ \ \forall t_{-i} \in T_{-i}
\]
Since $t_i' \in T_i$ was arbitrary, this holds for all $t_i\in T_i$. Then define
$k_i:T_{-i}\to {\bf R}$ by
\[
k_i(t_{-i}) = p_i(0, t_{-i})- \tilde p_i(0, t_{-i})
\]
By the argument above, for all $t_i \in T_i$,  $p_i(t_i, t_{-i}) -  \tilde p_i(t_i, t_{-i}) = k_i(t_{-i})$ for all $t_{-i}\in T_{-i}$. Since $i\in I$ was arbitrary, the first claim follows. 

For the second claim, suppose type $t^*_i$ of buyer $i$ is robustly indifferent between these mechanisms, so 
\begin{equation*}
E_{\pi }\left[v_{i}(t_{i}^{\ast},t_{-i}) q_{i}(t_{i}^{\ast },t_{-i})-p_{i}(t_{i}^{\ast },t_{-i})\right] =E_{\pi }\left[
v_{i}(t_{i}^{\ast },t_{-i})q_{i}(t_{i}^{\ast },t_{-i})-\tilde{p}%
_{i}(t_{i}^{\ast },t_{-i})\right] \ \ \forall \pi \in \Pi(t_{i}^{\ast })
\end{equation*}%
This implies
\begin{equation*}
E_{\pi }\left[ p_{i}(t_{i}^{\ast },t_{-i})\right] =E_{\pi }\left[ \tilde{p}%
_{i}(t_{i}^{\ast },t_{-i})\right] \ \ \ \forall \pi \in \Pi(t_{i}^{\ast })
\end{equation*}%
Because $\Pi(t_{i}^{\ast })$ has full dimension, this implies $p_{i}(t_{i}^{\ast
},t_{-i})=\tilde{p}_{i}(t_{i}^{\ast },t_{-i})$ for all $t_{-i}\in T_{-i}$.
By the argument above, for any $t_i \in T_i$,
\begin{equation*}
0 = p_i(t_i^*, t_{-i}) - \tilde p_i(t_i^*, t_{-i}) = p_i(t_i, t_{-i}) - \tilde p_i(t_i, t_{-i}) =  k_i(t_{-i})   \ \ \ \forall t_{-i} \in T_{-i} 
\end{equation*}%
Hence $k_i(t_{-i}) = 0 $ for all $t_{-i}\in T_{-i}$, and for all $t_i\in T_i$, $p_{i}(t_{i},t_{-i})=\tilde{p}_{i}(t_{i},t_{-i})$  for all $t_{-i}\in T_{-i}$.
\end{proof}

Theorem \ref{theorem:expost_revenue_equivalence} implies that the introduction of robustness to 
misspecification of beliefs can have a significant impact on the set of feasible auction mechanisms. 
The set of all interim incentive compatible mechanisms in the standard auction model with
independent types contains all equilibria of all standard auction formats, such as first-price, 
second-price, and all pay. Many of these mechanisms use ex post monotone allocation rules. 
Thus in many of these auction formats, the introduction of a small amount of misspecification can  
mean that only dominant strategy mechanisms satisfy robust incentive compatibility and remain 
feasible. In particular, there exists at most one mechanism $( q,p) $ for any given assignment rule $q$
that satisfies robust incentive compatibility and leaves at least one type of each buyer indifferent 
between participating and earning zero ex post utility.

\section{\label{section:GeneralModel}The General Model}

In this section we return to the single agent setting and extend the quasilinear model to allow for general outcomes, as well as general interdependence in utility over types, outcomes, and states. As before, the agent has privately known type $t\in T=[0,1]$. We consider the problem of designing a
mechanism to implement an abstract outcome chosen from a set $O$, where $O$ is a topological space.  As we do throughout the paper, we restrict attention here to direct mechanisms in which agents do not randomize reports. As in section 2, this is essentially without loss of generality.\footnote{We give a version of the revelation principle for the general setting in the appendix. As in the quasilinear case in section 2, ruling out randomized reports is essentially without loss of generality, under requisite additional regularity conditions, in particular if $\phi$ is Borel measurable and $u$ is bounded.} In addition, as in section \ref{section:Quasilinear}, mechanisms depends on an exogenous state $s$ drawn from a compact metric space $S$, which is publicly verifiable ex post.  Thus we consider mechanisms described by a function $\phi
:T\times S\rightarrow O$ that specifies an outcome $\phi( \theta
,s) \in O$ for any reported type $\theta \in T $ and any realized state 
$s\in S$. We assume $\phi(t, \cdot)$ is Borel measurable for each $t\in T$.  As shown in the previous section, this framework is easily
modified to allow for many agents.

The agent's payoff function is $u:O\times T\times S\rightarrow \mathbf{R}$.
Thus if the agent reports $\theta $ while her true type is $t$, her ex post
utility when the realized state is $s$ is $u(\phi (\theta ,s),t,s)$. 

We impose standard regularity conditions on the payoff function. 

\begin{assumption}
\label{Assumption:Utility_is_differentiable} The payoff function $u:O\times
T\times S\rightarrow \mathbf{R}$ is differentiable with respect to $t$, and $%
u_{2}(o,t,s):=\frac{\partial u}{\partial t}(o,t,s)$ is non-negative and bounded. In addition, $u(\cdot, t, \cdot)$ is Borel measurable for each $t\in T$. 
\end{assumption}

This assumption does not rule out the possibility that $u_{2}$
is zero on some interval. Thus two distinct types $t$ and $%
t^{\prime }$ can have the same payoff function $u(\cdot ,t,\cdot )=u(\cdot
,t^{\prime },\cdot )$ but differ in their beliefs, for example.

For this general framework, we start by recording the standard notions of
ex post and interim incentive compatibility.

\begin{definition}
\label{definition:EPIC}A mechanism $\phi :T\times S\rightarrow O\,$is \emph{%
ex post incentive compatible} if for each $t, \theta\in T$,
\begin{equation*}
u(\phi (t,s),t,s)\geq u(\phi (\theta ,s),t,s)  \ \
\forall s\in S
\end{equation*}
\end{definition}

\begin{definition}
\label{definition:BIC}A mechanism $\phi :T\times S\rightarrow O$ is \emph{%
interim incentive compatible} given $\{ \pi(t)\in \Delta(S):t\in T\}$ if for each $t, \theta \in
T$,
\begin{equation*}
E_{\pi (t)}\left[ u( \phi ( t,s) ,t,s) \right] \geq E_{\pi (t)}\left[ u( \phi ( \theta ,s) ,t,s) \right]
\end{equation*}
\end{definition}

As in the previous sections, we are interested in a stronger version of interim incentive compatibility that reflects robustness to possible misspecification of beliefs $\{\pi
(t):t\in T\}$. Again we model robustness by considering a set $\Pi(t)\subseteq \Delta(S)$ for each $t\in T$ and requiring that interim
incentive compatibility holds for each element of this set.

\begin{definition}
\label{definition:OIC}A mechanism $\phi :T\times S\rightarrow O$ is \emph{%
robust incentive compatible} for beliefs $\{ \Pi(t) \subseteq \Delta(S): t\in T\}$ if for all $t, \theta\in T$,
\begin{equation*}
E_{\pi }\left[ u( \phi ( t,s) ,t,s) \right] \geq E_{\pi }\left[  u( \phi ( \theta ,s) ,t,s) \right] \ \ \forall \pi \in \Pi (
t)
\end{equation*}
\end{definition}

As in the quasilinear case, these three notions of incentive compatibility are naturally nested. An ex post incentive compatible mechanism is also robust and interim  incentive compatible, while a robust incentive compatible mechanism is interim incentive compatible (provided $\pi(t) \in \Pi(t)$ for each $t$). Similarly, when $\Pi(t)$ is a singleton for each $t$, robust incentive compatibility reduces to interim incentive compatibility. At the other extreme, if $\Pi(t) = \Delta(S)$ for each $t$, then robust incentive compatibility is equivalent to ex post incentive compatibility. That is, a version of Lemma 1 carries over. We record this here for completeness, and omit the proof which mimics the proof of Lemma 1. 

\begin{lemma}
\label{lemma: full ignorance general} If $\Pi (t)=\Delta (S)$ for each $t\in T$,
then a mechanism $\phi$ is robust incentive compatible for beliefs $\{\Pi(t): t\in T\}$ if and only if it
is ex post incentive compatible.
\end{lemma}

Our main result in this section is the analogue of Theorem 3. When beliefs are fully overlapping, robust
incentive compatibility again generates significant  restrictions on the set of feasible
mechanisms. In particular, any robust incentive compatible mechanism must
satisfy an appropriately defined ex post envelope condition.

\begin{definition}
\label{definition:EPEC}A mechanism $\phi :T\times S\rightarrow O$ satisfies
the \emph{ex post envelope condition} if for each $t',t''\in T$,
\begin{equation*}
u(\phi (t'',s),t'',s)-u(\phi (t',s),t',s)=\int_{t'}^{t''}u_{2}(\phi (t ,s),t ,s) dt \ \ \ \forall s\in S
\end{equation*}
\end{definition}

As in the previous sections, this is an ex post version of revenue equivalence for this general environment. In particular, ex post utility differences across types are pinned down in this formula. 

All the ingredients are in place and we can now state the main result of this section.

\begin{theorem}
\label{theorem:OIC implies EPEC} If Assumption \ref%
{Assumption:Utility_is_differentiable} holds and beliefs $\{\Pi (t) \subseteq \Delta(S):t\in T\}$ are 
fully overlapping, then any robust incentive compatible mechanism must
satisfy the ex post envelope condition.
\end{theorem}

\begin{proof}
Let $\phi$ be a robust incentive compatible mechanism. Fix $t_0\in T$ and a neighborhood $N(t_0)$ of $t_0$ such that $\bigcap\limits_{t\in N(t_0)} \Pi(t)$ has full dimension. Fix $\pi\in \bigcap\limits_{t\in N(t_0)} \Pi(t)$. For each $t'\in N(t_0)$ set 
\[
w(t') = \max_{t\in T} E_\pi \left[ u(\phi(t,s), t', s) \right]
\]
Since $\phi$ is robust incentive compatible and $\pi\in \Pi(t')$ for each $t'\in N(t_0)$, $w$ is well defined, and $w(t') = E_\pi \left( u(\phi(t',s), t',s) \right)$. Then $u$ is uniformly Lipschitz continuous in $t'$, so $E_\pi(u(\phi(t,s), t',s))$ is Lipschitz continuous in $t'$ for each $t\in T$. 
This implies $w$ is Lipschitz continuous, and thus absolutely continuous and differentiable almost everywhere. By a version of the envelope theorem for Lipschitz functions due to Clarke (\cite{Clarke}, Theorem 2.8.6), for every $t'$ at which $w$ is differentiable, 
\[
w'(t') = E_\pi\left[ u_2(\phi(t',s),t',s)\right]
\]
Then for any $t',t'' \in N(t_0)$, 
\[
w(t'') - w(t') = \int_{t'}^{t''} E_\pi\left[ u_2(\phi(t,s),t,s)\right] dt
\]
by the fundamental theorem of calculus. 
Then for any $t',t''\in N(t_0)$, 
\begin{eqnarray*}
w(t'') - w(t') &=& E_\pi\left[ u(\phi(t'',s), t'',s)\right] - E_\pi \left[ u(\phi(t',s), t',s) \right]\\
&=& \int_{t'}^{t''} E_\pi \left[ u_2(\phi(t,s), t,s) \right] dt
\end{eqnarray*}
By using Tonelli's theorem, this implies
\[
E_\pi\left[ u(\phi(t'',s), t'',s)- u(\phi(t',s), t',s) \right] =  E_\pi \left[ \int_{t'}^{t''} u_2(\phi(t,s), t,s) dt \right] 
\]
Since $\pi\in \bigcap\limits_{t\in N(t_0)} \Pi(t)$ was arbitrary, 
\[
E_\pi\left[ u(\phi(t'',s), t'',s)- u(\phi(t',s), t',s) \right] =  E_\pi \left[ \int_{t'}^{t''} u_2(\phi(t,s), t,s) dt \right] \ \ \forall \pi\in \bigcap\limits_{t\in N(t_0)} \Pi(t)
\]
Since $\bigcap\limits_{t\in N(t_0)} \Pi(t)$ has full dimension, this implies
\[
u(\phi(t'',s), t'',s)- u(\phi(t',s), t',s)  =  \int_{t'}^{t''} u_2(\phi(t,s), t,s) dt  \ \ \forall s\in S  \tag{$\star\star\star$}
\]
Then $t', t''\in N(t_0)$ were arbitrary, so $(\star\star\star)$ holds for all $t', t''\in N(t_0)$. Since $t_0\in T$ was arbitrary, for every $t\in T$ there exists a neighborhood $N(t)$ of $t$ such that $(\star\star\star)$ holds for all $t', t''\in N(t)$. Now the result follows from the compactness and connectedness of $T$, as in the proof of Theorem 3. 
\end{proof}

As in the quasilinear case, we note that the weaker condition that  beliefs $\{\overline{\mbox{co}}(\Pi(t)): t\in T\}$ are fully overlapping is sufficient to imply that any robust incentive compatible mechanism for the beliefs $\{\Pi(t): t\in T\}$ must satisfy the ex post envelope condition. We omit the statement of this analogue of Corollary 1. 

Next, we show that fully overlapping beliefs and a monotonicity
condition appropriate for this general environment can collapse the set of
robust incentive compatible mechanisms to those that are ex post incentive
compatible. To that end, we begin by defining an appropriate ex post
monotonicity condition.

\begin{definition}
\label{definition:EPM}A mechanism $\phi :T\times S\rightarrow O$ satisfies 
\emph{ex post monotone type sensitivity} if for each $t \in T$, for $t', t'' \in T$, $t''\geq t' 
\Rightarrow u_{2}(\phi (t'',s),t,s)\geq u_{2}(\phi (t',s),t,s)$ for all $s\in S$.
\end{definition}

This condition reduces to standard ex post monotonicity conditions in the simplified environments of sections \ref{section:Quasilinear} and 3. In the case of quasilinear utility and private values, ex post monotone type sensitivity is equivalent to weak monotonicity as defined in 
\cite{Biketal06}. They provide a characterization of
ex post incentive compatibility in terms of weak monotonicity and a richness condition on the type space. Similarly, \cite{Rochet87} establishes an
equivalence between cyclical monotonicity and ex post incentive
compatibility. Ex post monotone type sensitivity is weaker than cyclical monotonicity, and is not sufficient for such an equivalence. Under the additional condition of fully overlapping beliefs, however, ex post monotone type sensitivity implies that any robust incentive compatible mechanism must be ex post incentive compatible.

\begin{theorem}
\label{theorem:OIC and EPIC} Suppose that Assumption \ref%
{Assumption:Utility_is_differentiable} holds and that beliefs $\{\Pi (t) \subseteq \Delta(S):t\in T\}$
are fully overlapping. Then any robust incentive compatible mechanism that
satisfies ex post monotone type sensitivity is ex post incentive compatible.
\end{theorem}

\begin{proof}
By Theorem \ref{theorem:OIC implies EPEC}, any robust incentive compatible mechanism 
satisfies the ex post envelope condition. Thus we must show that if a mechanism satisfies the ex post envelope condition and ex post monotone type sensitivity, then it must
be ex post incentive compatible.

To that end, fix $t''\in T$ and $s\in S$. First consider $t'\in T$ with $t''\geq t'$. Then 
\[
u(\phi(t'',s), t'',s) - u(\phi(t',s), t',s) = \int_{t'}^{t''} u_2(\phi(t,s), t,s) dt \geq \int_{t'}^{t''} u_2(\phi(t',s), t,s) dt
\]
where the first equality follows from the ex post envelope condition, and the inequality follows from ex post monotone type sensitivity. 
By the fundamental theorem of calculus, 
\[
u(\phi(t'',s), t'',s) - u(\phi(t',s), t',s)  \geq \int_{t'}^{t''} u_2(\phi(t',s), t,s) dt = u(\phi(t',s), t'',s) - u(\phi(t', s), t', s)
\]
Thus 
\[
u(\phi(t'',s), t'',s) - u(\phi(t',s), t',s)  \geq u(\phi(t',s), t'',s) - u(\phi(t', s), t', s)
\]
or equivalently, $u(\phi(t'',s), t'',s) \geq u(\phi(t',s), t'',s)$. 

Now suppose $t'\geq t''$. Then again using the ex post envelope condition and ex post monotone type sensitivity, 
\begin{eqnarray*}
u(\phi(t'',s), t'',s) - u(\phi(t',s), t',s) &=& \int_{t'}^{t''} u_2(\phi(t,s), t,s) dt \\
 &=& - \int_{t''}^{t'} u_2(\phi(t,s), t,s) dt \\
&\geq& -\int_{t''}^{t'} u_2(\phi(t',s), t,s) dt
\end{eqnarray*}
By the fundamental theorem of calculus,
\begin{eqnarray*}
u(\phi(t'',s), t'',s) - u(\phi(t',s), t',s)  &\geq& -\int_{t''}^{t'} u_2(\phi(t',s), t,s) dt \\
&=& u(\phi(t',s), t'',s) - u(\phi(t', s), t', s)
\end{eqnarray*}
So again $u(\phi(t'',s), t'',s) \geq u(\phi(t',s), t'',s)$. Since $t'' \in T$ and $s\in S$ were arbitrary, for all $t'' \in T$,
\[
u(\phi(t'',s), t'',s) \geq u(\phi(t',s), t'',s) \ \ \forall t'\in T \mbox{ and } \forall s\in S
\]
Thus $\phi$ is ex post incentive compatible. 
\end{proof}

Results analogous to those at the end of section \ref{section:Quasilinear} can be derived to explore the extent to which robust incentive compatibility embeds some notion of monotonicity. In particular, results paralleling Theorem 5 are straightforward and are left to the reader.

\section{\label{section:Preferences}Knightian Uncertainty and Robustness}

In this section, we discuss a behavioral foundation for our notion of robustness, in which beliefs $\{\Pi (t) \subseteq \Delta(S) :t\in T\}$ are derived endogenously and reflect perceptions of ambiguity. Knightian decision theory, developed in Bewley (1986), provides the link between our notion of robustness and ambiguity. This model gives an alternative justification for robust incentive compatibility as robustness to the presence of ambiguity or Knightian uncertainty. This yields a parallel interpretation of our main results as showing that ambiguity can have a significant impact on  mechanism design by limiting the set of feasible mechanisms.  

In Bewley (1986), ambiguity is modeled by incomplete preferences.\footnote{Bewley's original paper has been published as \cite{Bewley02}. See also Ghirardato et al. (2003), Gilboa et al. (2010), and Girotto and Holzer (2005).} Bewley (1986) axiomatizes incomplete preference relations that can be represented by a family of
subjective expected utility functions, showing that a
 preference relation that is not necessarily complete, but satisfies other standard axioms of subjective expected utility, can be represented using a von Neumann-Morgenstern utility index and a set of probability distributions, with preference corresponding
to unanimous ranking according to all elements of this set. The preference relation is complete if and only if this set is a singleton, in which case the  standard subjective expected utility representation obtains. Thus the decision maker perceives ambiguity if and only if the preference relation is incomplete, and  both the amount of ambiguity perceived and the
degree of incompleteness of the preference relation are measured by the size of the corresponding set of probabilities.\footnote{See  \cite{Ghirardato-Maccheroni-Marinacci} or \cite{Rigotti-Shannon} for precise results along these lines.}

This model is easily adapted to our setup. An agent's privately known type $t$ characterizes his preference relation $\succsim_t$ over state-contingent outcomes, which are Borel measurable functions from $S$ to $O$. The agent has a utility function $u:O\times T\times
S\rightarrow \mathbf{R}$, where for each $t\in T$, $u(\cdot, t, \cdot)$ is Borel measurable.  Each
type $t\in T$ is associated with a set of beliefs $\Pi(t)
\subseteq \Delta(S) $.\footnote{In the representation in Bewley (1986), the set of probabilities is compact and convex. Although we have allowed beliefs to be general subsets of $\Delta(S)$ throughout the paper, it would be without loss of generality to assume that  the sets $\Pi(t)$ are closed and convex, as we noted in section 2.1.} For Borel measurable functions $x, y:S\to O$, 
\[
x\succsim_t y \ \ \mbox{ if and only if } \ E_\pi[u(x(s), t, s)] \geq E_\pi[u(y(s), t,s)] \ \ \forall \pi\in \Pi(t)
\]

A mechanism $\phi:T\times S \to O$ maps report-state pairs into outcomes. Given the agent's type and preference relation over state-contingent outcomes, the mechanism $\phi$ then induces a preference relation over reports. Given the mechanism $\phi$, an agent of type $t$ prefers report $\theta$ to report $\theta'$ if and only if
\[
E_\pi[u(\phi(\theta, s), t, s)] \geq E_\pi[u(\phi(\theta', s), t,s)] \ \ \forall \pi\in \Pi(t)
\] 
A mechanism is then robust incentive compatible if and only if each type prefers truthful reporting over any other reporting strategy in this mechanism. In this interpretation of the model, robust incentive compatible mechanisms are those that provide agents with no incentive to misreport, as in a standard Bayesian environment.   

Robust incentive compatibility can also be viewed as reflecting robustness to the agent's attitude toward ambiguity, or to objective versus subjective uncertainty in the terminology of \cite{Ghirardato-Maccheroni-Marinacci}, and  \cite{Gilboa-Maccheroni-Marinacci-Schmeidler}. In this interpretation, the designer may identify a set of possible beliefs for the agent, but does not know the precise representation of the agent's preferences. Many different models of ambiguity are consistent with this framework, including maxmin expected utility, Choquet expected utility, variational preferences,  the smooth ambiguity model and other models of second-order priors. This idea is similar to the separation between beliefs and ambiguity attitudes in \cite{Ghirardato-Maccheroni-Marinacci}, \cite{GMMS}, and in \cite%
{Rigotti-Shannon-Strzalecki08}. Robust incentive compatibility then corresponds to the requirement that the mechanism be robust to the designer's lack of detailed knowledge regarding how ambiguity is perceived by the agent.

Similarly, this can be understood as a model of collective choice, in which the agent represents a group of decision makers instead of a single decision maker. The agents in this group share a common utility function and private information but might have different beliefs, and make choices based on unanimous rankings. This is a common interpretation of Bewley preferences; related models of social preferences are emphasized and studied recently in \cite{Danan-Gajdos-Hill-Tallon}, \cite{Alon-Gayer}, \cite{Gilboa-Samuelson-Schmeidler}, and \cite{Brunnermeier-Simsek-Xiong}, for example. In this setting, robust incentive compatible mechanisms are the mechanisms in which the group always prefers to report truthfully.

\section{\label{section:FullExtraction}Robustness and Surplus  Extraction}

In this section we consider a robust version of the surplus extraction problem. The designer can typically extract all, or virtually all, information rents  whenever agents' beliefs are correlated with their private information in standard Bayesian models. This has been a central puzzle in mechanism design, and has motivated significant attention to developing foundations for mechanisms less sensitive to fine  details of the environment, particularly agents' beliefs. The main results of the previous sections show that robust incentive compatibility can provide such a foundation in many settings when beliefs are fully overlapping. In particular, by showing that only ex post incentive compatible mechanisms are feasible for the designer in many environments, these results typically rule out full extraction which instead relies on stochastic payments leveraging the correlation between  agents' beliefs and information. While fully overlapping beliefs arise naturally in a variety of settings, as we showed above, other natural conditions can instead lead to beliefs that are not fully overlapping, including parametric restrictions or moment conditions. The results in this section provide some insight into the general scope for surplus extraction in such settings, as well as limits on the designer.

We first give a generalization of the classic results of Cr{\'e}mer and McLean (1985, 1988) and \cite{McAfeeReny92}. We show that the designer can achieve virtual extraction whenever agents' beliefs satisfy a natural set-valued analogue of the convex independence conditions of Cr{\' e}mer-McLean and McAfee-Reny. Virtual extraction frequently fails in our setting, in contrast with the standard Bayesian model in which full or virtual extraction is generically possible in many contexts. We then study limits on the designer's ability to extract surplus. When virtual extraction fails, we show that the designer can be restricted to simple mechanisms. When beliefs are fully overlapping, robust incentive compatibility limits the designer to offering a single contract, and additional natural conditions can make a deterministic contract optimal for the designer. These results suggest that the designer might often face a tradeoff between eliciting information and generating revenue. We also give some results on partial extraction in this spirit.
 
To isolate general properties leading to surplus extraction and provide a result applicable in many different problems, we follow McAfee and Reny (1992) in giving a reduced form description of the surplus extraction problem. In a prior, unmodeled stage, agents play a game that leaves them with some information rents as a function of their private information. Private information is summarized by the type $t\in T$. As in previous sections, we let $T = [0,1]$ be the set of types.\footnote{For all of the results in this section, it suffices that $T$ is a compact, convex metric space.}  The current stage also has an exogenous source of uncertainty, summarized by the set of states $S$, on which contract payments can depend, as in the previous sections. For some applications it is natural to take $S=T^n$ for some $n$ as in section 3, although as in previous sections and in McAfee and Reny (1992), we take  $S$ to be an arbitrary compact metric space. 

To each type $t\in T$ is then associated a value $v(t)\in {\bf R}$, representing the rents from the prior stage, and a nonempty set of beliefs $\Pi(t) \subseteq \Delta (S)$. Throughout this section we maintain the assumption that $v:T\to {\bf R}$ is continuous. Our main results on surplus extraction also assume that $\Pi(t)$ is convex and norm compact for each $t\in T$, and that the correspondence $\Pi:T\to 2^{\Delta(S)}$ is norm continuous. This provides a natural extension of the setting in McAfee and Reny (1992), in which each type $t$ is associated with a unique belief $\pi(t) \in \Delta(S)$ and the map $\pi:T\to \Delta(S)$ is assumed to be norm continuous. We use the following notation throughout this section. If $x(t) \in C(S)$ for each $t\in T$ we write $x(t)(s) = x(t, s)$ for each $s\in S$. Similarly, for $x\in C(T\times S)$ and $t\in T$, we write $x(t) \in C(S)$ for the function such that $x(t) (s) = x(t,s)$ for each $s\in S$. We use $r\in {\bf R}$ interchangeably for the constant $r$ and the function $r{\bf 1}_S$, where ${\bf 1}_S$ denotes the identity on $S$. 

We start by formalizing notions of full extraction and virtual extraction in this setting. 

\bigskip
\begin{definition}
\emph{Full extraction} holds if, for each given $v:T\to {\bf R}$, there exists a collection $\{c(t) \in C(S) : t\in T\}$ such that for each $t\in T$:
\[ 
v(t) - E_{\pi}[c(t,s)] = 0 \ \ \ \forall \pi\in \Pi(t)
\]
and
\[
v(t) -E_{\pi}[c(t',s)] \leq 0 \ \ \ \forall t'\not= t, \ \forall \pi\in \Pi(t)
\]
\emph{Virtual extraction} holds if, for each given $v:T\to {\bf R}$ and for each $\varepsilon >0$, there exists a collection $\{c_\varepsilon(t)\in C(S): t\in T\}$ such that for each $t\in T$:
\[
0\leq v(t) - E_{\pi}[c_\varepsilon(t,s)] \leq \varepsilon\ \ \ \forall \pi\in \Pi(t)
\]
and
\[
0\leq \sup_{t'\in T} \lbrace v(t) - E_{\pi}[c_\varepsilon (t',s)] \rbrace \leq \varepsilon \ \ \ \forall \pi\in \Pi(t)
\]
\end{definition}

\bigskip

These notions of full and virtual extraction are natural extensions of the standard notions that account for uncertainty in agents' beliefs.  We take these to be the robust versions of full and virtual extraction in this setting. Both reflect the idea that the designer offers agents a menu of stochastic contracts from which they choose, based on their expected costs. For the case of full extraction, the requirement that the contract $c(t)$ leave type $t$ with zero expected surplus for each belief $\pi\in \Pi(t)$ might seem too strong, and the weaker requirement that $v(t) -E_{\pi}[c(t,s)] \geq 0$ for all $\pi\in \Pi(t)$, with $v(t) = E_{\pi}[c(t,s)]$ for some $\pi\in \Pi(t)$, might seem more natural. We show below in Lemma 5 that this weaker notion of full extraction is in fact equivalent to full extraction as defined above in this setting. Nonetheless, this can be a strong condition. For example, if $\Pi(t)$ has full dimension for some $t$, then $c(t)=v(t)$ and full extraction will not be possible in general. 

As defined above, full extraction or virtual extraction might require the designer to offer an infinite menu of contracts. In the case of virtual extraction, even in the standard case in which $\Pi(t)$ is a singleton for each $t$, such a menu need not have an expected cost minimizing element for all agents. By allowing for an infinite menu of contracts, this might also appear to be a weaker notion of virtual extraction than considered by McAfee and Reny (1992). In the model of McAfee and Reny (1992), each type  has a single belief $\pi(t) \in \Delta(S)$; their main result then gives conditions on the set $\{ \pi(t) : t\in T\}$ such that for each $v:T\to {\bf R}$ and for each $\varepsilon >0$, there is a finite menu $\{ c_1, \ldots ,c_n\} \subseteq C(S)$ such that for each $t\in T$, 
\[
0\leq \max_{j=1,\ldots ,n} \lbrace v(t) -E_{\pi(t)}[c_j(s)] \rbrace \leq \varepsilon 
\]
We note that whenever virtual extraction holds in our setting (using the definition above), then it is always possible to find a finite menu of contracts that would achieve the same bounds on surplus. We record this observation below; the proof is in the appendix together with other proofs omitted from this section.

\begin{lemma}
Suppose $\Pi(t)$ is convex and norm compact for each $t\in T$, and the correspondence $\Pi:T\to 2^{\Delta(S)}$ is norm continuous. 
If virtual extraction holds, then virtual extraction can be achieved with a finite menu of contracts. That is, given $v:T\to {\bf R}$, for each $\varepsilon >0$ there exists a finite menu $\{ c_1,\ldots ,c_n\}\subseteq C(S)$ such that for each $t\in T$, there exists $c_i\in \{c_1,\ldots ,c_n\}$ such that 
\[
0\leq v(t) -E_{\pi}[c_i(s)]  \leq \varepsilon  \ \ \forall \pi\in \Pi(t)
\]
and for all $c_j\in \{ c_1,\ldots ,c_n\}$,
\[
v(t) -E_{\pi}[c_j(s)]  \leq \varepsilon  \ \ \forall \pi\in \Pi(t)
\]
\end{lemma}

Our main result of this section shows that virtual extraction holds when beliefs satisfy the following condition, which we call probabilistic independence. This is a natural set-valued analogue of the conditions identified by Cr{\' e}mer and McLean (1985, 1988) and McAfee and Reny (1992). 

\begin{definition}
Types satisfy \emph{probabilistic independence} if for any $\mu\in \Delta(T)$,  given a type $t_0\in T$ and an element $\pi^1(t_0) \in \Pi(t_0)$, and given  a  measurable selection $\pi^2:T\to \Delta(S)$ with $\pi^2(t) \in \Pi(t)$ for every $t\in T$,  
\[
 \pi^1(t_0) =  \int \pi^2(t) d\mu  \Rightarrow \mu = \delta_{t_0}
\]
\end{definition}

\bigskip

\noindent {\bf Remark: } When $\Pi(t)$ is a singleton for each $t$, this model reduces to that of McAfee and Reny (1992). In this case, probabilistic independence is the condition McAfee and Reny (1992) identify as sufficient for virtual surplus extraction. When $\Pi(t) = \{ \pi(t)\}$ for each $t\in T$, then probabilistic independence requires that for each $t_0\in T$,
\[
 \pi(t_0) =  \int \pi(t) d\mu \mbox{ for some } \mu\in \Delta(T)  \Rightarrow \mu = \delta_{t_0}
\]
Similarly, when $T$ is finite, probabilistic independence is a natural set-valued analogue of the {\it convex independence} condition identified by Cr{\' e}mer and McLean (1985, 1988). In this  case probabilistic independence requires that given a type $t_0\in T$ and an element $\pi^1(t_0) \in \Pi(t_0)$,  and given any selection with $\pi^2(t) \in \Pi(t)$ for every $t\in T$,  
\[
 \pi^1(t_0) =  \sum_{t\in T} \mu_t \pi^2(t) \mbox{ for some } \mu\in \Delta(T)  \Rightarrow \mu_{t_0} = 1
\]

The main result in this section shows that under probabilistic independence, virtual extraction is possible in this model. That is, a robust analogue of the virtual extraction result of McAfee and Reny (1992) holds. As with the original result of McAfee and Reny (1992), which is nested in ours, probabilistic independence is not sufficient to guarantee full extraction in general with infinitely many types (see \cite{Lopomo-Rigotti-Shannon-detect}). Nonetheless, in order to explain the result and the strategy of the proof, we start by considering the problem of full extraction. 

We start by casting the problem of full or virtual extraction in slightly  stronger terms. We will see that while this results in stronger conditions, probabilistic independence nonetheless guarantees that this stronger form of virtual extraction holds. Rather than looking for a collection of contracts $\{ c(t) \in C(S): t\in T\}$, we add the requirement that the contracts also be jointly continuous in types and states, and thus consider the existence of a schedule of contracts $c\in C(T\times S)$ such that for each $t\in T$:
\begin{eqnarray*}
v(t) - E_{\pi}[c(t,s)] &=& 0 \ \ \ \forall \pi\in \Pi(t)\\
v(t) - E_{\pi}[c(t',s)] &\leq & 0 \ \ \ \forall \pi\in \Pi(t), \ \forall t'\not= t
\end{eqnarray*}

First we note that this is equivalent to the seemingly weaker condition {\it weak full extraction}, requiring for each $t\in T$:
\begin{eqnarray*}
v(t) - E_{\pi}[c(t,s)] &\geq& 0 \ \ \ \forall \pi\in \Pi(t)\\
v(t) - E_{\pi}[c(t',s)] &\leq & 0 \ \ \ \forall \pi\in \Pi(t), \ \forall t'\not= t
\end{eqnarray*}

We establish this equivalence in the lemma below. 

\begin{lemma}
Suppose $\Pi(t)$ is convex and norm compact for each $t\in T$, and the correspondence $\Pi:T\to 2^{\Delta(S)}$ is norm continuous. 
Then for each $v:T\to {\bf R}$,  $c\in C(T\times S)$ satisfies full extraction if and only if $c$ satisfies weak full extraction.
\end{lemma}

From Lemma 5, it is enough to consider the relaxed problem of weak full extraction. Now write 
\[
c(t) = v(t) + z(t) 
\]
where $z\in C(T\times S)$ and we use $v(t) \in {\bf R}$ interchangeably with $v(t) {\bf 1}_S$, where ${\bf 1}_S$ denotes the identity on $S$. Note that any $c$ can be written this way for appropriate choice of $z$. Then $c$ satisfies full extraction if and only if for each $t\in T$:
\begin{eqnarray*}
E_{\pi}[z(t,s)] &\leq & 0 \ \ \ \forall \pi\in \Pi(t)\\
E_{\pi}[z(t',s)] &\geq & v(t) - v(t') \ \ \ \forall \pi\in \Pi(t), \ \forall t'\not= t
\end{eqnarray*}
This follows from observing that for each $t$ and $\pi\in \Pi(t)$,
\[
v(t) - E_{\pi}[c(t,s)] = v(t) - v(t) - E_{\pi}[z(t,s)] = -E_{\pi}[z(t,s)] 
\]
and
\[
v(t) - E_{\pi}[c(t',s)] = v(t) - v(t') - E_{\pi}[z(t',s)] 
\]
Thus we will consider the existence of $z\in C(T\times S)$ such that each $t\in T$:
\begin{eqnarray*}
E_{\pi}[z(t,s)] &\leq & 0 \ \ \ \forall \pi\in \Pi(t)\\
E_{\pi}[z(t',s)] &\geq & v(t) - v(t') \ \ \ \forall \pi\in \Pi(t), \ \forall t'\not= t
\end{eqnarray*}

To that end, let $f:T\times C(T\times S) \to {\bf R}$ be given by 
\[
f(t,z) = \max_{\pi\in \Pi(t)} E_{\pi}[z(t,s)]
\]
and for each $t\in T$, let $f_t:C(T\times S) \to {\bf R}$ be given by $f_t(z) = f(t,z)$. Similarly, let $g:T\times T\times C(T\times S)\to {\bf R}$ be given by 
\[
g_{(t,t')}(z) := g(t,t',z) = \min_{\pi\in \Pi(t)} E_{\pi}[z(t',s)]
\]
Then to show that full extraction is possible, it suffices to show that there exists $z\in C(T\times S)$ such that 
\begin{eqnarray*}
f(t,z) & \leq & 0 \ \ \forall t\in T\\
g(t,t',z) &\geq & v(t) - v(t') \ \ \forall t'\not= t, \ \forall t\in T
\end{eqnarray*}
Note that for each $t\in T$, $f_t : C(T\times S) \to {\bf R}$ is convex, and for every $t,t' \in T$, $g_{(t,t')}: C(T\times S)\to {\bf R}$ is concave. In addition, $f(0) = g(0)=0$.

Now consider the problem 
\begin{equation}
\begin{aligned}
p^* =& \underset{c\in {\bf R}, \; z\in C(T\times S)}{\text{inf}}
& & c \\
& \text{subject to}
& & f(t,z) \leq c \ \ \ \forall t\in T \\
&&& v(t) - v(t') -g(t,t',z) \leq c \ \ \ \forall t,t' \in T
\end{aligned}
\tag{vse}
\end{equation}

Note that if the optimal value $p^*$ of this problem is less than or equal to zero, then at least virtual surplus extraction is possible. We establish this in the next lemma.

\begin{lemma}
Suppose $\Pi(t)$ is convex and norm compact for each $t\in T$, and the correspondence $\Pi:T\to 2^{\Delta(S)}$ is norm continuous. 
If $p^*\leq 0$, then virtual surplus extraction holds. If $p^*<0$, or if $p^*=0$ and is $p^*$ is obtained in (vse), then full extraction holds. 
\end{lemma}

Thus to show that virtual surplus extraction is possible, it suffices to show that $p^*\not> 0$. We establish this by considering the dual of the optimization problem (vse), and making use of duality to argue that these problems have the same value. The heart of the proof is then to show that this common value cannot be positive under probabilistic independence. 

\begin{theorem}
Suppose $\Pi(t)$ is convex and norm compact for each $t\in T$, and the correspondence $\Pi:T\to 2^{\Delta(S)}$ is norm continuous. 
If types satisfy probabilistic independence, then virtual extraction holds. 
\end{theorem}

We close this section with some results on the limits to virtual extraction when probabilistic independence fails. We identify a natural condition under which incentive compatibility limits the variation in contracts that the designer can offer, in some cases limiting the designer to a single contract. 

We start with a general definition of incentive compatibility in this setting. The menu $C= \{c(t) \in C(S): t\in T\}$ is \emph{robust incentive compatible} if for each $t\in T$: 
\[
v(t) - E_{\pi}[c(t,s)] \geq v(t) - E_{\pi}[c(t',s)] \ \  \forall \pi\in \Pi(t), \ \forall t'\not= t
\]
Similarly, say $C$ is \emph{robustly individually rational} if for each $t\in T$:
\[
v(t) - E_{\pi}[c(t,s)] \geq 0 \ \ \forall \pi\in \Pi(t)
\]

Robust incentive compatibility can impose significant restrictions on feasible menus of contracts, depending on the extent to which different types have common beliefs, since it requires that two contracts $c(t)$ and $c(t')$ must have the same expected value according to any belief $\pi$ that is shared by types $t$ and $t'$. When the set of such shared beliefs is sufficiently rich, this observation forces $c(t)$ and $c(t')$ to be the same. When beliefs are fully overlapping, the only robust incentive compatible menus contain a single contract. We collect these results below.

\begin{theorem}
Suppose $C = \{ c(t)\in C(S) : t\in T\}$ is robust incentive compatible. 
\begin{enumerate}[(i)]
  \item If $\Pi(t) \cap \Pi(t')$ has full dimension, then $c(t) = c(t')$. 
  \item If beliefs $\{ \Pi(t) \subseteq \Delta(S): t\in T\}$ are fully overlapping then $c(t) = c(t')$ for all $t, t'\in T$. 
  \end{enumerate}
\end{theorem}
\begin{proof}
For (i), fix $t, t'\in T$ with $t\not= t'$. Since $C$ is robust incentive compatible, 
\[
v(t) - E_{\pi}[c(t',s)] \leq v(t) - E_{\pi}[c(t,s)] \ \ \forall \pi\in \Pi(t)
\]
and 
\[
v(t') - E_{\pi}[c(t,s)] \leq v(t') - E_{\pi}[c(t',s)] \ \ \forall \pi\in \Pi(t')
\]
Thus
\[
E_{\pi}[c(t',s)] \geq E_{\pi}[c(t,s)] \ \ \forall \pi\in \Pi(t)
\]
and
\[
E_{\pi}[c(t,s)] \geq E_{\pi}[c(t',s)] \ \ \forall \pi\in \Pi(t')
\]
Putting these together, 
\[
E_{\pi}[c(t,s) - c(t',s)] = 0 \ \ \forall \pi\in \Pi(t)\cap \Pi(t')
\]
Since $\Pi(t)\cap \Pi(t')$ has full dimension, this implies $c(t,s) = c(t',s)$ for all $s\in S$, that is, $c(t) = c(t')$.

For (ii), since beliefs are fully overlapping, for each $t\in T$ there exists a neighborhood $N(t)$ such that $\cap_{t'\in N(t)} \Pi(t')$ has full dimension. In particular, $\Pi(t') \cap \Pi(t) $ has full dimension for all $t'\in N(t)$.  Thus by (i), $c(t') = c(t)$ for all $t'\in N(t)$. 

Now we claim that $c(t) = c(t')$ for all $t, t'\in T$, from the compactness and connectedness of $T$. To see this, note that $\{ N(t): t\in T\}$ is an open cover of $T$ and $T$ is compact, so there exist $t_1, \ldots ,t_n$ such that $T\subseteq \cup_i N(t_i)$. If $T\subseteq N(t_1)$, then we are done, as $c(t) = c(t_1)$ for all $t\in N(t_1)$. Otherwise, if $T\not\subseteq N(t_1)$, then there must exist some $t_i\not= t_1$ such that $N(t_1) \cap N(t_i)\not= \emptyset$; without loss of generality take $t_i =t_2$. This follows from the fact that $T$ is connected and $T\subseteq \cup_i N(t_i)$; otherwise $N(t_1) \cap ( \cup_{i=2}^n N(t_i)) = \emptyset$ and $T\subseteq N(t_1) \cup (\cup_{i=2}^n N(t_i))$, with $T\cap N(t_1) \not= \emptyset$ and $T\cap (\cup_{i=2}^n N(t_i)) \not=\emptyset$,  where $N(t_1)$ and $\cup_{i=2}^n N(t_i)$ are open, which would contradict the connectedness of $T$. Since there exists some $\bar t\in N(t_1) \cap N(t_2)$, $c(t_1) = c(\bar t) = c(t_2)$. Thus $c(t) = c(t_1)$ for all $t\in N(t_1) \cup N(t_2)$. Similarly, if $T\subseteq N(t_1) \cup N(t_2)$ we are done. If $T\not\subseteq N(t_1) \cup N(t_2)$, then there exists $t_i\in \{ t_3,\ldots t_n\}$ such that $(N(t_1) \cup N(t_2)) \cap N(t_i) \not= \emptyset$; without loss of generality take $t_i = t_3$. Otherwise, $(N(t_1) \cup N(t_2)) \cap (\cup_{i=3}^n N(t_i)) = \emptyset$ and $T\subseteq (N(t_1) \cup N(t_2)) \cup (\cup_{i=3}^n N(t_i))$, with $T\cap \left(N(t_1)\cup N(t_2)\right) \not= \emptyset$ and $T\cap \left(\cup_{i=3}^n N(t_i)\right) \not=\emptyset$, where $N(t_1) \cup N(t_2)$ and $\cup_{i=3}^n N(t_i)$ are open, again contradicting the connectedness of $T$. Then, as above, this implies $c(t_1) = c(t_2) = c(t_3)$. Thus $c(t) = c(t_1)$ for all $t\in N(t_1)\cup N(t_2) \cup N(t_3)$. By repeating this argument, since $n$ is finite, this implies $c(t_1) = c(t_i)$ for all $i=2,\ldots ,n$.  Thus $c(t) = c(t_1)$ for all $t\in T$, from which (ii) follows.
\end{proof}

Thus even if the designer chooses to offer a smaller menu of contracts, for example by forgoing participation for some types with lower values to focus on extracting more surplus from types with higher values, these results imply that in a robust incentive compatible menu the designer must always offer the same contract to any pair of types sharing a set of beliefs of full dimension, and when beliefs are fully overlapping the designer can offer at most one contract. 

Our main example illustrates these results. If $\pi:T\to \Delta(S)$ is norm continuous and $\{ \pi(t): t\in T\}$ satisfies probabilistic independence, then the designer can achieve virtual extraction, which follows from the result of McAfee and Reny (1992). Suppose in addition that $\pi(t)$ and $\pi(t')$ are mutually absolutely continuous for all $t,t'\in T$, $\frac{d\pi(t')}{d\pi(t)}$ is continuous in $t$ for each $t'\in T$, and the family $\{ \frac{d\pi(t')}{d\pi(t)} : t\in T\}$ is equicontinuous. Then by Theorem 2, the beliefs $\{ \Pi_\varepsilon(t): t\in T\}$ are fully overlapping for any $\varepsilon >0$,  where $\Pi_\varepsilon(t) = \{(1-\varepsilon)\pi(t) + \varepsilon \pi : \pi\in  \Delta(S)\}$. Thus virtual extraction holds for $\varepsilon = 0$, but for any $\varepsilon >0$, any robust incentive compatible menu for beliefs $\{\Pi_\varepsilon (t) : t\in T\}$ limits the designer to a single contract for this setting. 

With some additional information about the designer's beliefs $\pi(d) \in \Delta(S)$, these results lead to precise  predictions regarding the designer's optimal menu of contracts. Although these results require additional restrictions that we do not focus on otherwise, they might have some independent interest. Thus we record these results next. 

\begin{theorem}
Suppose beliefs $\{ \Pi(t) \subseteq \Delta(S): t\in T\}$ are fully overlapping. Let $t_1\in T$ satisfy $v(t_1) = \min_{t\in T} v(t)$. If $\pi(d) \in \Pi(t_1)$, then the deterministic contract $v(t_1)$ maximizes the designer's expected revenue among all  menus $C = \{ c(t) \in C(S): t\in T\}$ that satisfy robust incentive compatibility and robust individual rationality. \end{theorem}
\begin{proof}
Let $C = \{ c(t) \in C(S): t\in T\}$ satisfy robust incentive compatibility and robust individual rationality. By part (ii) of Theorem 12, $c(t) = c(t')= c(t_1)$ for all $t, t'\in T$. Since $C$ satisfies robust individual rationality, 
\[
v(t_1) - E_{\pi}[c(t_1,s)] \geq 0 \ \ \ \forall \pi\in \Pi(t_1)
\]
Thus 
\[
E_{\pi}[c(t_1,s)] \leq v(t_1) \ \ \ \forall \pi\in \Pi(t_1)
\]
If $\pi(d) \in \Pi(t_1)$, this implies $E_{\pi(d)}[c(t_1,s)] \leq v(t_1)$. Now note that the menu in which $c(t) = v(t_1)$ for each $t\in T$ satisfies robust incentive compatibility and robust individual rationality. Thus the deterministic contract $v(t_1)$ maximizes the designer's expected revenue among all such menus. 
\end{proof}

Note that similar conclusions follow if the designer has a set of beliefs $\Pi(d) \subseteq \Delta(S)$. In that case, the deterministic contract $v(t_1)$ is optimal for the designer among all menus that satisfy robust incentive compatibility and robust individual rationality, provided $\Pi(d)\subseteq \Pi(t_1)$. If instead the designer uses a maxmin criterion, then the deterministic contract $v(t_1)$ maximizes the designer's minimum expected revenue among all menus that satisfy robust incentive compatibility and robust individual rationality as long as $\Pi(d) \cap \Pi(t_1) \not= \emptyset$.

We close this section with some results on the possibilities for surplus extraction if the designer focuses only on a subset of types. 

\begin{theorem}
Suppose $S$ is finite. 
\begin{itemize}
  \item[(i)] Suppose $\Pi(t) \subseteq \Delta(S)$ is closed for each $t\in T$.  If $\{ t_1, \ldots , t_n\} \subseteq T$ and $\{ \Pi(t_i): i=1,\ldots, n\}$ satisfies probabilistic independence, then weak full extraction holds for $\{t_1, \ldots ,t_n\}$. 
  \item[(ii)] Suppose $\pi:T\to \Delta(S)$ and $\{ \pi(t): t\in T\}$ satisfies probabilistic independence. For every $\{t_1,\ldots ,t_n\} \subseteq T$ there exists $\varepsilon >0$ such that weak full extraction holds for $\{t_1, \ldots ,t_n\}$ with respect to beliefs $\{\Pi_\varepsilon(t_i): i=1,\ldots, n\}$.   
\end{itemize}   
\end{theorem}
\begin{proof}
For (i), since $\{\Pi(t_i): i=1,\ldots ,n\}$ satisfies probabilistic independence, for each $i=1,\ldots, n$:
\[
\Pi(t_i) \cap \mbox{co} (\cup_{j\not= i}\Pi(t_j)) = \emptyset
\]
By assumption $\Pi(t_j) \subseteq \Delta(S) \subseteq {\bf R}^S$ is closed for each $j$ and $S$ is finite, so $\mbox{co} (\cup_{j\not= i}\Pi(t_j)) = \overline{\mbox{co}}  (\cup_{j\not= i}\Pi(t_j))$. Thus for each $i=1,\ldots ,n$, 
\[
\Pi(t_i) \cap \overline{\mbox{co}} (\cup_{j\not= i}\Pi(t_j)) = \emptyset
\]
The result now  follows from Theorem 2 in \cite{Lopomo-Rigotti-Shannon-finite}. 

For (ii), fix $\{t_1,\ldots ,t_n\}\subseteq T$. Since $\{\pi(t): t\in T\}$ satisfies probabilistic independence, so does $\{\pi(t_i): i=1,\ldots ,n\}$. Thus for each $i=1,\ldots ,n$: 
\[
\pi(t_i) \notin \overline{\mbox{co}} \{ \pi(t_j): j\not= i\}
\]
Then there exists $\varepsilon >0$ such that for each $i=1,\ldots ,n$:
\[
\Pi_\varepsilon(t_i) \cap \overline{\mbox{co}} (\cup_{j\not= i}\Pi_\varepsilon(t_j)) = \emptyset
\]
Now the result follows from (i). 
\end{proof}

\section{\label{section:Conclusions}Conclusion}

We have studied mechanism design problems robust to misspecification of beliefs. Our main results focused on the implications of robust incentive compatibility in a general class of environments, motivated either by the designer's lack of information about agents' beliefs, or by agents' perceptions of ambiguity. Our main results show that robust incentive compatibility can often put strong restrictions on the set of feasible mechanisms, limiting the designer to ex post incentive compatible mechanisms under common conditions. 

There are a number of promising directions for further work. In many environments it is natural to consider limits on the set of possible beliefs, including parametric restrictions or moment conditions. Such restrictions can rule out fully overlapping beliefs, yet still impose significant structure on the set of feasible mechanisms. Similarly, we have focused primarily on characterizing the set of feasible mechanisms under robust incentive compatibility. Understanding optimal mechanisms for various revenue or welfare goals under robust incentive compatibility is another central question. A number of recent papers give important results along both of these lines for particular problems and under natural restrictions on beliefs. Examples include Bergemann and Schlag (2008, 2011) and \cite{Carrasco-Luz-Kos-Messner-Monteiro},  on the problem of a monopoly seller facing a single buyer, and \cite{Neeman}, \cite{Kocyigit-Iyengar-Kuhn-Wiesemann-20}, and \cite{Suzdaltsev}, on the problem of a seller facing multiple bidders for a single indivisible item. Many open questions along similar lines remain and would be important directions for future work.


\vfill\eject

\section{Appendix}

\subsection{Proofs from Section 6}

This appendix contains proofs for the results in section 6. 

We start with some basic notation that will be maintained throughout this section. For a compact metric space $B$, $C(B)$ is the space of continuous real-valued functions on $B$, and ${\cal M}(B)$ is the space of finite signed Borel measures on $B$. For $x\in C(B)$ and $\eta \in {\cal M}(B)$, we write $x\cdot \eta = \eta \cdot x$ for the bilinear form $\langle x, \eta \rangle = \langle \eta, x\rangle$, that is, 
\[
\eta \cdot x = x\cdot \eta = \langle x, \eta \rangle = \langle \eta, x \rangle = \int x(b) \, \eta(db) = \int x(b) \, d\eta
\]
In particular, throughout this section of the appendix, for $x\in C(B)$ and $\pi\in \Delta(B)$, we use the notation
\[
\pi \cdot x = E_\pi [x(b)] = \int x(b) \, \pi(db) = \int x(b) \, d\pi
\]

Before giving the proofs of the results in section 6, we first collect several lemmas and additional results that will be used in the arguments below. 

We start by considering an alternative, sequential notion of probabilistic independence, which is weaker than probabilistic independence in general, and is useful in some arguments. 

\begin{definition}
Types satisfy \emph{sequential probabilistic independence} if for any $\mu\in \Delta(T)$, any  sequence $\{ \pi^1_n(t_0)\}$ for some $t_0\in T$, and any sequence of measurable selections $\{ \pi^2_n(t), t\in T\}$ such that $\pi^1_n(t_0)\in \Pi(t_0)$, $\pi^2_n(t) \in \Pi(t)$ for each $n$ and each $t\in T$, 
\[
\left\Vert \pi^1_n(t_0) - \int \pi^2_n(t) \mu(dt) \right\Vert \to 0 \Rightarrow \mu = \delta_{t_0}
\]
\end{definition}

\bigskip

Sequential probabilistic independence implies probabilistic independence; this follows from the definitions by considering only constant sequences. The converse also holds provided $\Pi(t)$ is norm compact and convex for every $t\in T$. We establish this converse in the following lemma.

\bigskip

\begin{lemma}
If $\Pi(t)$ is convex and norm compact for each $t\in T$, then types satisfy probabilistic independence if and only if types satisfy sequential probabilistic independence. 
\end{lemma}
\begin{proof}
As noted above, sequential probabilistic independence implies probabilistic independence by definition. For the converse, fix $\mu\in \Delta(T)$ and $t_0\in T$. Let $\{\pi^1_n(t_0)\}$ be a sequence with $\pi^1_n(t_0) \in \Pi(t_0)$ for each $n$ and $\{ \pi^2_n(t), t\in T\}$ be a sequence of measurable selections $\pi^2_n:T\to \Delta(S)$ with $\pi^2_n(t) \in \Pi(t)$ for each $n$ and $t\in T$ such that 
\[
\left\Vert \pi^1_n(t_0) - \int \pi^2_n(t) \mu(dt) \right\Vert \to 0 
\]
We must show $\mu = \delta_{t_0}$. To that end, let
\begin{multline*}
\int \Pi(t) \mu(dt) := \\ \{ \gamma \in \Delta(S) : \gamma = \int \pi(t) \mu(dt) \mbox{ for some msble selection } \pi(t) \in \Pi(t) \ \forall t\in T\}
\end{multline*}
Then note that by definition $\int \Pi(t) \mu(dt)$ is the Aumann integral of $\Pi$ with respect to $\mu$. Since $\Pi(t)$ is norm compact and convex for each $t\in T$, $\int \Pi(t) \mu(dt)$ is also the Debreu integral of $\Pi$ with respect to $\mu$, and thus is norm compact and convex (see \cite{Debreu} and \cite{Byrne}). Thus there are subsequences $\{ \pi^1_{n_k}(t_0)\}, \{ \pi^2_{n_k}(t), t\in T\}$, without loss of generality using the same indexes, an element $\pi^1(t_0) \in \Pi(t_0)$, and a measurable selection $\pi^2:T\to \Delta(S)$ with $\pi^2(t) \in \Pi(t)$ for each $t\in T$ such that 
\[
\pi^1_{n_k}(t_0) \to \pi^1(t_0) \mbox{ and } \int \pi^2_{n_k}(t) \mu(dt) \to \int \pi^2(t) \mu(dt)
\]
Then, as $\left\Vert \pi^1_{n_k}(t_0) - \int \pi^2_{n_k}(t) \mu(dt) \right\Vert \to 0 $, 
\[
\left\Vert \pi^1(t_0) - \int \pi^2(t) \mu(dt) \right\Vert = 0 
\]
that is,
\[
 \pi^1(t_0) = \int \pi^2(t) \mu(dt) 
\]
Since types satisfy probabilistic independence, this implies $\mu = \delta_{t_0}$. 
\end{proof}

The next lemma collects some basic properties of functions used in the arguments in section 6.

\begin{lemma}
\begin{itemize}
  \item[(i)] Let $h:T\times \Delta(S) \times C(T\times S)\to {\bf R}$ be given by $h(t,\pi, z) = \pi\cdot z(t)$. Then $h$ is continuous. 
  \item[(ii)]  Let $x\in C(S)$. Suppose $\Pi(t)$ is norm compact for each $t\in T$, and the correspondence $\Pi:T\to 2^{\Delta(S)}$ is norm continuous. Then the functions $t\mapsto \max_{\pi\in \Pi(t)} \pi\cdot x$ and $t\mapsto \min_{\pi\in \Pi(t)} \pi\cdot x$ are continuous. 
\end{itemize}
\end{lemma}
\begin{proof}
For (i), fix $(t,\pi,z) \in T\times \Delta(S) \times C(T\times S)$, and let $(t_n, \pi_n, z_n)\to (t, \pi,z)$. Then 
\begin{eqnarray*}
\Vert \pi_n\cdot z_n(t_n)-\pi \cdot z(t) \Vert &=& \Vert (\pi_n - \pi ) \cdot z_n(t_n) + \pi\cdot (z_n(t_n) -z(t) ) \Vert\\
&\leq& \Vert  (\pi_n - \pi ) \cdot z_n(t_n)\Vert + \Vert \pi\cdot (z_n(t_n) -z(t) ) \Vert\\
&\leq& \Vert \pi_n - \pi \Vert \Vert z_n(t_n)\Vert + \Vert \pi\cdot (z_n(t_n) -z(t) ) \Vert\\
\end{eqnarray*}
Since $z_n\to z$, $\{ z_n(t_n)\} $ and $\{ z_n(t_n) - z(t) \}$ are bounded, and $z_n(t_n)\to z(t)$ pointwise. Then $\Vert \pi_n - \pi \Vert \Vert z_n(t_n)\Vert\to 0$, since $\pi_n\to \pi$ in norm, and $\pi\cdot (z_n(t_n) -z(t) ) \to 0$ by the bounded convergence theorem. Thus $h(t_n, \pi_n, z_n) \to h(t,\pi, z)$.

For (ii), let $x\in C(S)$. By assumption, $\Pi(t)$ is norm compact for each $t\in T$ and the correspondence $t\mapsto \Pi(t)$ is norm continuous. Then the claim follows from Berge's Theorem. 
\end{proof}

Because these results might be of independent interest, we include here the derivation of the version of Ekeland's Variational Principle that we use in the proof of Theorem 11 below.  We start with some preliminary definitions and results, including the classic version of Ekeland's Variational Principle, and an extension due to Borwein, from which the version we use here follows quickly. See also \cite{Lopomo-Rigotti-Shannon-detect}.

\begin{definition}
For $x\in \mbox{ dom } f$ and $\varepsilon >0$, the \emph{$\varepsilon$-subdifferential} of $f$ at $x$, denoted $\partial_\varepsilon f(x)$ is 
\[
\partial_\varepsilon f(x) = \{ x^*\in X^*: f(y) \geq f(x) + \langle x^*, y-x\rangle -\varepsilon \ \ \forall y\in X\}
\]
\end{definition}

\noindent {\bf Note: } For any $x\in \mbox{ dom } f$ and any $\varepsilon >0$,
\begin{itemize}
  \item $\partial_\varepsilon f(x)$ is closed and convex for every $x$
  \item $\inf f \leq f(x_\varepsilon ) \leq \inf f + \varepsilon \iff 0\in \partial_\varepsilon f(x_\varepsilon)$
  \item for $\varepsilon = 0$, $\partial_0 f(x) = \partial f(x)$
  \item if $f$ is convex and lsc, then $\partial_\varepsilon f(x) \not=\emptyset$ for every $x\in \mbox{ dom } f$
  \end{itemize}  

Next we state the classic version of Ekeland's Variational Principle (see \cite{Ekeland}). 

\begin{theorem}
{\rm \bf (Ekeland's Variational Principle)} Let $V$ be a complete metric space and $F:V\to {\bf R}\cup \{ +\infty\}$ be a proper, lsc function such that $\inf F > -\infty$. Let $\varepsilon >0$ and $\beta >0$. For every $u\in V$ such that 
\[
\inf F \leq F(u) \leq \inf F + \varepsilon
\]
there exists $v\in V$ such that 
\begin{itemize}
  \item[(i)] $F(v) \leq F(u)$
  \item[(ii)] $d(u,v) \leq \beta$
  \item[(iii)] $F(u) \geq F(v) - \frac{\varepsilon}{\beta} d(v,w) \ \ \ \forall w\not= v$   
\end{itemize}
If in addition $F$ is convex, then 
\begin{itemize}
  \item[(iv)] $v$ can be chosen such that there exists $g\in \partial F(v)$ such that $\Vert g \Vert \leq \frac{\varepsilon}{\beta}$  
\end{itemize}
\end{theorem}

More precise approximations can be given for convex functions, as shown by \cite{Borwein}.  

\begin{theorem}
{\rm \bf (Borwein, 1982, Theorem 1)} Let $X$ be a Banach space and  $f:X\to {\bf R} \cup \{ +\infty\}$ be a proper, convex, lsc function. Let $\varepsilon >0$ and $k\geq 0$ be given. Let 
\[
x_0^*\in \partial_\varepsilon f(x_0)
\]
Then there exist $x_\varepsilon$ and $x_\varepsilon^*$ such that
\[
x_\varepsilon^* \in \partial f(x_\varepsilon)
\]
and such that
\begin{eqnarray*}
\Vert x_\varepsilon - x_0\Vert &\leq & \sqrt{\varepsilon}\\
\vert f(x_\varepsilon) - f(x_0) \vert &\leq& \sqrt{\varepsilon} (\sqrt{\varepsilon} + \frac{1}{k})\\
\Vert x_\varepsilon^* - x_0^* \Vert &\leq & \sqrt{\varepsilon} (1+ k\Vert x_0^*\Vert )\\
\vert x_\varepsilon^*(h) - x_0^*(h)\vert &\leq & \sqrt{\varepsilon} ( \Vert h \Vert + k\vert x_0^*(h)\vert)\\
x_\varepsilon^* &\in& \partial_{2\varepsilon}f(x_0)
\end{eqnarray*}
\end{theorem}

Putting these two results together yields the following. 

\begin{lemma}
{\bf (\cite{Lopomo-Rigotti-Shannon-detect})} Let $X$ be a Banach space and $f:X \to {\bf R}$ be a proper, convex, lsc function such that $\inf f > -\infty$. Then there exists a sequence $\{ x_n\}$ in $X$ such that $f(x_n) \to \inf f$ and $d(0, \partial f(x_n)) \to 0$, i.e., there exists $\{ g_n\}$ such that $g_n \in \partial f(x_n)$ for each $n$ and $\Vert g_n \Vert \to 0$. 
\end{lemma}



\bigskip

\begin{lemma}
Let $B$ be a compact metric space. Let $f:B\times C(B) \to {\bf R}$ be continuous, and  for each $b\in B$, let $f_b:C(B)\to {\bf R}$ be given by $f_b(x) = f(b,x)$. Suppose $f_b$ is convex for each $b\in B$. Let $h:C(B)\to {\bf R}$ be given by 
\[
h(x) = \int f_b(x) \mu(db)
\]
where $\mu \in {\cal M}(B)$ and $\mu\geq 0$. Then $h$ is convex, and 
\[
\partial h(x) = \int \partial f_b(x) \mu(db)
\]
That is, for each $\gamma \in \partial h(x)$ there is a mapping $b\mapsto \gamma_b$ such that $\gamma_b \in \partial f_b(x)$ for $\mu$-$\mbox{ a.e } b\in B$ and 
\[
\gamma\cdot y = \int \gamma_b \cdot y \ \mu(db)
\]
for any measurable $y$. 
\end{lemma}
\begin{proof}
Since $B$ is a compact metric space, $C(B)$ is separable. The result then follows from \cite{Ioffe-Levin}; see also \cite{Clarke} Theorem 2.7.2 and discussion on pp. 76-77.
\end{proof}

\begin{lemma}
Suppose $\Pi(t)$ is convex and norm compact for each $t\in T$, and the correspondence $\Pi:T\to 2^{\Delta(S)}$ is norm continuous. 
Let $f:T\times C(T\times S) \to {\bf R}$ be given by 
\[
f(t,z) = \max_{\pi\in \Pi(t)} \pi \cdot z(t)
\]
and for each $t\in T$, let $f_t:C(T\times S) \to {\bf R}$ be given by $f_t(z) = f(t,z)$. 
Then $f$ is continuous and $f_t$ is convex for each $t\in T$. For each $t\in T$,  if 
$\gamma \in \partial f_t(z)$ then $\gamma \in {\cal M}(T\times S)$ is a measure such that for any measurable function $y$,
\[
 \gamma \cdot y = \pi \cdot y(t) \mbox{ for some } \pi \in \Pi(t)
\]
\end{lemma}
\begin{proof}
First, $f$ is continuous, by Lemma 8(i) and Berge's Theorem. By construction, $f_t$ is convex for each $t\in T$. Then for each $t\in T$ and each $\pi\in \Pi(t)$, let 
$h_t^\pi:C(T\times S ) \to {\bf R}$ be given by
\[
h_t^\pi(z) = \pi\cdot z(t)
\] 
Then note that $h_t^\pi$ is linear and continuous, and 
\[
f_t(z) = \max\limits_{\pi\in \Pi(t)} h_t^\pi(z)
\]
Given $t\in T$ and $\pi\in \Pi(t)$, $\partial h_t^\pi(z)$ is the measure $\gamma_t^\pi\in {\cal M}(T\times S)$ such that 
\[
\gamma_t^\pi\cdot y = \pi\cdot y(t) \ \ \mbox{ for $y$ measurable function }
\]
Then 
\[
\partial f_t(z) = w^*-\mbox{cl} \left( \mbox{co} \bigcup\limits_{\pi\in \Pi^z(t)} \partial h_t^\pi(z) \right)
\]
where $\Pi^z(t) = \{ \pi\in \Pi(t): f_t(z) = \pi\cdot z(t) = h_t^\pi(z) \}$ and  $w^*-\mbox{cl}(A)$ denotes the weak$^*$ closure of the set $A$. This follows from standard results on subgradient calculus for pointwise maxima; e.g., see Theorem 2.4.18 in \cite{Zalinescu-02}. 

For any $z\in C(T\times S)$, 
\begin{eqnarray*}
\mbox{co} \bigcup\limits_{\pi\in \Pi^z(t)} \partial h_t^\pi(z) &=& \mbox{ co} \{ \gamma\in {\cal M}(T\times S): \gamma\cdot y = \pi\cdot y(t) , \ \pi\in \Pi^z(t)\}\\
&\subseteq & \mbox{ co} \{ \gamma\in {\cal M}(T\times S): \gamma\cdot y = \pi\cdot y(t) , \ \pi\in \Pi(t)\}\\
&=& \{ \gamma\in {\cal M}(T\times S): \gamma\cdot y = \pi\cdot y(t) , \ \pi\in \Pi(t)\}
\end{eqnarray*}
Then $\Pi(t)$ is $w^*$-closed (since it is norm closed and convex), so 
\[
\partial f_t(z) = w^*-\mbox{cl} \left( \mbox{co} \bigcup\limits_{\pi\in \Pi^z(t)} \partial h_t^\pi(z)\right) \subseteq  \{ \gamma\in {\cal M}(T\times S): \gamma\cdot y = \pi\cdot y(t) , \ \pi\in \Pi(t)\}
\] 
The result follows. 
\end{proof}

\begin{lemma}
Suppose $\Pi(t)$ is convex and norm compact for each $t\in T$, and the correspondence $\Pi:T\to 2^{\Delta(S)}$ is norm continuous. 
Let $g:T\times T\times C(T\times S)\to {\bf R}$ be given by 
\[
g_{(t',t)}(z) := g(t',t,z) = \min_{\pi\in \Pi(t')} \pi\cdot z(t)
\]
Then $g$ is continuous, and $g_{(t',t)}$ is concave for each $t,t'\in T$. For each $t,t'\in T$, if $\gamma \in \partial g_{(t',t)}(z)$ then $\gamma \in {\cal M}(T\times S)$ is a measure such that for any measurable function $y$,
\[
 \gamma \cdot y = \pi \cdot y(t) \mbox{ for some } \pi \in \Pi(t')
\]
\end{lemma}
\begin{proof}
This follows from arguments analogous to those used in the proof of Lemma 11. 
\end{proof}

We now turn to proofs of the results in section 6. 

\noindent {\it Proof of Lemma 4.} 
Let $v:T\to {\bf R}$ be given, and fix $\varepsilon >0$. Choose $c_\varepsilon \in C(T\times S)$ such that for each $t\in T$:
\[
0\leq v(t) - \pi \cdot c_\varepsilon (t) \leq \varepsilon \ \ \ \forall \pi\in \Pi(t) 
\]
and
\[
0\leq \sup_{t'\in T} \lbrace v(t) - \pi \cdot c_\varepsilon (t') \rbrace \leq \varepsilon \ \ \ \forall \pi\in \Pi(t)
\]
So for each $t\in T$, 
\[
0\leq v(t) -\max_{\pi\in \Pi(t)} \pi\cdot c_\varepsilon(t) \leq v(t) -\min_{\pi\in \Pi(t)} \pi\cdot c_\varepsilon(t) \leq \varepsilon
\]
By Lemma 8(ii), the function $t'\mapsto \max_{\pi\in \Pi(t')} \pi\cdot c_\varepsilon(t) $ is continuous. Thus for each $t\in T$ there exists $\delta_t>0$ such that $t'\in B_{\delta_t}(t) \Rightarrow$
\[
-\frac{\varepsilon}{2} \leq v(t') - \max_{\pi\in \Pi(t')} \pi\cdot c_\varepsilon(t) \leq v(t') - \pi\cdot c_\varepsilon(t) \leq \varepsilon \ \ \ \forall \pi\in \Pi(t')
\]
Thus for $t'\in B_{\delta_t}(t)$,
\[
-\frac{\varepsilon}{2} \leq v(t') - \pi\cdot c_\varepsilon(t) \leq \varepsilon \ \ \ \forall \pi\in \Pi(t')
\]
Since $T$ is compact and $\{B_{\delta_t}(t) : t\in T\}$ is an open cover of $T$, there exists $t_1,\ldots ,t_n$ such that $T\subseteq \cup_i B_{\delta_{t_i}}(t_i)$. Then for each $i$, set $c_i = c_\varepsilon(t_i) + \varepsilon$. For each $t\in T$ there exists $i$ such that $t\in B_{\delta_{t_i}}(t_i)$. Thus 
\[
0\leq v(t) -\pi \cdot c_i \leq 2\varepsilon \ \ \ \forall \pi\in \Pi(t)
\]
and for each $j=1,\ldots ,n$, 
\[
v(t) -\pi \cdot c_j \leq 2\varepsilon \ \ \ \forall \pi\in \Pi(t)
\]
The claim follows. 
\hfill \qed

\noindent {\it Proof of Lemma 5.} 
Full extraction clearly implies weak full extraction. To see that these are equivalent, fix $v:T\to {\bf R}$ and  $t\in T$. Then choose a sequence $t_n\to t$ with $t_n \not= t$ for each $n$; this is possible by the connectedness of $T$.  Suppose $c\in C(T\times S)$ satisfies weak full extraction. Fix $\pi \in \Pi(t)$. Then 
\[ 
v(t) - \pi \cdot c(t) \geq 0 
\]
and for each $n$, since $t_n \not= t$,  
\[
v(t) -\pi \cdot c(t_n) \leq 0 
\]
Since $c\in C(T\times S)$, $\{ c(t_n)\}$ is bounded and $c(t_n) \to c(t)$ pointwise. By the bounded convergence theorem, $\pi\cdot c(t_n) \to \pi\cdot c(t)$. Thus
\[
v(t) - \pi \cdot c(t) \leq 0
\]
Thus 
\[
v(t) -\pi \cdot c(t) = 0
\]
Since $\pi \in \Pi(t)$ and $t\in T$ were arbitrary, the equivalence follows.
\hfill \qed

\noindent {\it Proof of Lemma 6.} 
To see this, first suppose $p^*<0$. Then there must exist $z\in C(T\times S)$ such that 
\[
f(t, z) \leq 0 \ \ \ \forall t\in T
\]
and
\[
v(t) - v(t') - g(t,t',z) \leq 0 \ \ \ \forall t,t'\in T
\]
In this case, full extraction holds, and thus a fortiori, virtual extraction holds as well. Similarly, if $p^* = 0$, then either $p^*$ is realized, in which case again there must exist such a $z\in C(T\times S)$ as above so that full extraction holds, or if $p^*$ is not realized, then for each $\varepsilon >0$ there exists $z_\varepsilon \in C(T\times S)$ such that 
\[
f(t, z_\varepsilon) \leq \varepsilon \ \ \ \forall t\in T
\]
and
\[
v(t) - v(t') - g(t,t',z_\varepsilon) \leq \varepsilon \ \ \ \forall t,t'\in T
\]
Now note that using continuity of $z_\varepsilon$, $g$, and $v$, this implies
\[
\min_{\pi \in \Pi(t) } \pi\cdot z_\varepsilon (t) = g(t,t,z_\varepsilon) \geq -\varepsilon \ \ \ \forall t\in T
\]
Thus
\[
-\varepsilon \leq \pi(t) \cdot z_\varepsilon (t) \leq \varepsilon \ \ \forall \pi\in \Pi(t), \ \ \forall t\in T
\]
Now set 
\[
z:= z_\varepsilon - \varepsilon
\]
and for each $t\in T$, set the contract $c(t)$ to be 
\[
c(t) = v(t) + z(t)
\]
Then $c\in C(T\times S)$, and for each $t\in T$ and each $\pi \in \Pi(t)$, 
\begin{eqnarray*}
v(t) - \pi \cdot c(t) &=& v(t) - v(t) -\pi \cdot z(t) \\
&=& -\pi \cdot z(t) \\
&=& \varepsilon - \pi \cdot z_\varepsilon (t)
\end{eqnarray*}
and by the preceding argument,
\[
0 \leq \varepsilon - \pi \cdot z_\varepsilon (t)    \leq 2\varepsilon
\]
Thus for each $t\in T$, 
\[
0\leq v(t) -\pi \cdot c(t) \leq 2\varepsilon \ \ \ \forall \pi\in \Pi(t)
\]
Then fix $t\in T$ and $\pi\in \Pi(t)$, and consider $t'\not= t$. 
\begin{eqnarray*}
v(t) - \pi \cdot c(t') &=& v(t) - v(t') -\pi \cdot z(t') \\
&=& v(t) - v(t') -\pi \cdot z_\varepsilon (t') + \varepsilon\\
&\leq & v(t) -v(t') -g(t,t',z_\varepsilon) + \varepsilon\\
&\leq & 2\varepsilon
\end{eqnarray*}
Thus for all $t\in T$, 
\[
v(t) - \pi \cdot c(t') \leq 2\varepsilon \ \ \ \forall \pi\in \Pi(t), \ \ \forall t'\not= t
\]
So for all $t\in T$,
\[
0\leq \sup_{t'\in T} v(t) - \pi \cdot c(t') \leq 2\varepsilon \ \ \ \forall \pi\in \Pi(t)
\]
The result follows. 
\hfill \qed

\noindent {\it Proof of Theorem 11.}
By Lemma 6, to show that virtual surplus extraction is possible, it suffices to show that $p^* \not> 0$. To that end, note that the Lagrange dual function for the problem $(\mbox{vse})$ is 
\[
{\cal L}(\lambda, \nu) = \inf_{{ c\in {\bf R} }\atop {z\in C(T\times S)}} \left\lbrace c + \lambda \cdot (f(t,z) -c) + \nu \cdot (v(t) - v(t') -g(t,t',z) - c) \right\rbrace
\]
Let $d\in C(T\times T)$ be given by 
\[
d(t,t') = v(t) - v(t')
\]
Note that $d(t,t) = 0$ for all $t\in T$. 

Define 
\[
h(\lambda, \nu ) = \inf_{z\in C(T\times S)} \left\{ \lambda \cdot f(z) + \nu\cdot (d-g(z))\right\}
\]
Using this notation, we can rewrite the Lagrange dual function for $(\mbox{vse})$ as follows:
\[
{\cal L}(\lambda, \nu) = \left\lbrace 
\begin{array}{lr}
h(\lambda, \nu) & \text{ if } \int \lambda (dt) + \iint \nu(dt'\ dt) = 1\\
-\infty & \text{ otherwise }
\end{array}
\right.
\]
Thus the dual problem of $(\mbox{vse})$ is 
\begin{equation}
\begin{aligned}
d^* = & \underset{\lambda \in {\cal M}(T), \; \nu\in {\cal M}(T\times T)}{\text{sup}}
& & h(\lambda, \nu) \\
& \text{subject to}
& & (\lambda, \nu) \geq 0 \\
&&& \int \lambda (dt) + \iint \nu(dt'\ dt) = 1 
\end{aligned}
\tag{d-vse}
\end{equation}
Then note that Slater's condition holds for the original problem $(\mbox{vse})$. To see this, set $z=0$, so 
\[
f(t,z) = g(t,t',z) = 0 \ \ \ \forall t,t' \in T
\]
Then choose 
\[
\bar c > \sup_{t,t'\in T} v(t) - v(t') \geq 0
\]
For $(z,c) = (0, \bar c)$, 
\[
\sup_{t\in T} f(t,z) - c = -\bar c <0
\]
and 
\[
\sup_{t,t'\in T} v(t) - v(t') - g(t,t',z) -c = \sup_{t,t'\in T} v(t) - v(t') -\bar c <0
\]
Thus $p^*=d^*$ and in addition $d^*$ is obtained, where $p^*$ is the optimal value of $(\mbox{vse})$ and $d^*$ is the optimal value of $(\mbox{d-vse})$. 

Now it suffices to show that $p^* = d^* \not> 0$. To show this, suppose by way of contradiction that $p^* = d^* >0$. Since $d^*$ is obtained in $(\mbox{d-vse})$, there exists $(\lambda, \nu) \geq 0$ such that 
\[
d^* = h(\lambda, \nu) >0 \mbox{ and } \int \lambda(dt) + \iint \nu(dt'\ dt) =1
\]
Recall that, by definition, 
\[
h(\lambda, \nu) = \inf_{z\in C(T\times S)} \left( \lambda \cdot f(z) + \nu \cdot (d-g(z) ) \right)
\]
and $f(0) = g(0) =0$, which implies
\[
h(\lambda , \nu) \leq \nu \cdot d
\]
Since $h(\lambda, \nu) >0$, this implies $\nu\cdot d >0$. Thus $\nu\not= 0$. Since $\lambda, \nu \geq 0$, this implies $\nu>0$. 

Let $F:C(T\times S) \to {\bf R}$ be given by
\[
F(z) = \lambda \cdot f(z) + \nu\cdot (d-g(z))
\]
Note that $F$ is convex and continuous, and by definition, 
\[
h(\lambda, \nu) = \inf_{z\in C(T\times S)} F(z) >0
\]
In particular, this implies $\inf_{z\in C(T\times S)} F(z) \in {\bf R}$. By Ekeland's Variational Principle (see Lemma 9), there exists a sequence $\{ z_n\}$ and a sequence $\{ \gamma_n\}$ with $\gamma_n \in \partial F(z_n)$ for each $n$ such that
\[
F(z_n) = \lambda\cdot f(z_n) + \nu\cdot (d-g(z_n) ) \to \inf_{z\in C(T\times S)} F(z)= h(\lambda, \nu) >0
\]
and
\[
\Vert \gamma_n \Vert \to 0
\]
By Lemmas 10, 11, and 12, since $\gamma_n \in \partial F(z_n)$ for each $n$, there exist selections $\{ \pi_n(t), t\in T\}, \{ \pi^t_n(t'), t,t'\in T\}$ with $\pi_n(t)\in \Pi(t), \pi^t_n(t') \in \Pi(t')$ for each $n$ and each $t,t'\in T$, such that $\gamma_n$ is the measure for which 
\[
\gamma_n\cdot y = \int \pi_n(t) \cdot y(t)\ \lambda (dt) - \iint \pi^t_n(t') \cdot y(t)\ \nu (dt'\ dt)
\]
for any measurable function $y$. Since $\Vert \gamma_n \Vert \to 0$, $\gamma_n \cdot y \to 0$ for any such $y$. 

Now fix a measurable set $A\subseteq T$ and let $y$ be given by
\[
y(t) = \left\lbrace 
\begin{array}{lr}
0 & \text{ if } t\not\in A\\
{\bf 1}(S) & \text{ if } t\in A
\end{array}
\right.
\]
where ${\bf 1}(S)$ is the indicator of $S$. Then 
\begin{eqnarray*}
\gamma_n \cdot y &=& \int_A \pi_n(t) \cdot y(t)\ \lambda (dt) - \iint_{T\times A} \pi^t_n(t') \cdot y(t) \ \nu(dt'\ dt)\\
&=& \int_A \lambda(dt) - \iint_{T\times A} \nu(dt'\ dt)\\
&=& \lambda(A) - \nu(T\times A)
\end{eqnarray*}
Since $\gamma_n \cdot y \to 0$, this implies $\lambda(A) - \nu(T\times A)\to 0$, that is, $\lambda(A) = \nu(T\times A)$. Since $A$ was arbitrary, 
$\lambda(A) = \nu(T\times A)$ for each $A\subseteq T$. From this it follows first that $\lambda(T) = \nu(T\times T)$; since $\nu>0$, this implies $\lambda(T) = \nu(T\times T) >0$. Then without loss of generality, rescaling if necessary, take $\lambda(T) = \nu(T\times T) = 1$. Second, using disintegration of measures, this implies we can write
\[
\nu = \int \nu_t(dt') \lambda(dt)
\]
where $\nu_t$ is a measure on $T$, $\nu_t \geq 0$ and $\nu_t(T) = 1$ for each $t$ in the support of $\lambda$. 

For  each $n$ and each $t\in T$, let 
\[
\gamma_n(t) = \pi_n(t) - \int \pi^t_n(t') \nu_t(dt')
\]
Then for each $n$, $\gamma_n$ is the measure given by
\[
\gamma_n(E) = \int \gamma_n(t) (E_t) \lambda(dt) \ \ \ \ \ \forall E\subseteq T\times S
\]
where for $E\subseteq T\times S$, $E_t:= \{ s\in S: (t,s) \in E\}$. 

Then note that 
\begin{eqnarray*}
\Vert \gamma_n \Vert &=& \sup_E \vert \gamma_n(E) \vert \ \ \ \ \mbox{ by definition}\\
&=& \gamma_n^+(T\times S) + \gamma_n^-(T\times S) \ \ \ \ \mbox{ by definition }\\
&=& \int \left\lbrack \gamma_n^+(t) (S) + \gamma_n^-(t) (S) \right\rbrack \lambda(dt)\\
&=& \int \Vert \gamma_n(t)\Vert \lambda (dt) 
\end{eqnarray*}
Recall from above $\Vert \gamma_n \Vert =  \int \Vert \gamma_n(t)\Vert \lambda (dt)  \to 0$. Now set 
\[
\alpha(t)  = \liminf_n \Vert \gamma_n(t) \Vert \ \ \ \forall t\in T
\]
By Fatou's Lemma, 
\begin{eqnarray*}
0 \leq \int \alpha(t) \lambda (dt) &\leq& \liminf_n \int \Vert \gamma_n(t) \Vert \lambda (dt) \\
&=& 0  \ \ \ \ \mbox{ since } \int \Vert \gamma_n(t)\Vert  \lambda(dt) = \Vert \gamma_n\Vert \to 0
\end{eqnarray*}
Thus 
\[
\int \alpha(t) \lambda(dt ) = 0
\]
By definition, $\alpha(t) \geq 0$ for each $t\in T$, hence $\alpha(t) = 0$ for $\lambda-\mbox{a.e } t\in T$. 

Thus for $\lambda-\mbox{a.e } t\in T$, there is a subsequence $\{ \gamma_{n_k}(t)\}$ such that 
\[
\Vert \gamma_{n_k}(t) \Vert \to 0
\]
But for each $t\in T$,
\[
\Vert \gamma_{n_k}(t) \Vert = \Vert  \pi_{n_k}(t) - \int \pi^t_{n_k}(t') \nu_t(dt') \Vert
\]
where for each $t\in T$ and for each $n_k$, $\pi_{n_k}(t) \in \Pi(t)$, $\pi^t_{n_k}(t') \in \Pi(t')$ for each $t'\in T$, and $\nu_t \in \Delta(T)$. 
Thus by probabilistic independence and Lemma 7, $\nu_t = \delta_t$ for $\lambda-\mbox{a.e } t\in T$. 

But then
\begin{eqnarray*}
\nu\cdot d &=& \iint d(t',t) \nu(dt'\ dt) \\
&=& \iint d(t',t) \nu_t(dt') \lambda(dt)\\
&=& \int_{\mbox{supp } \lambda } d(t,t) \lambda (dt)\\
&=& 0  \ \ \ \ \mbox{ since } d(t,t) = 0 \mbox{ for all } t\in T
\end{eqnarray*}
This is a contradiction, as $\nu\cdot d >0$. Thus $p^* \leq 0$. 
\hfill \qed

\subsection{Direct Mechanisms and the Revelation Principle}

In this section we define abstract mechanisms and verify that a version of the revelation principle holds for our setting. We focus on the single agent case, but multiple agents are easily accommodated. Notation and basic setup are as in Section \ref{section:GeneralModel}. In particular, we take as given beliefs $\{\Pi(t)\subseteq \Delta(S): t\in T\}$. Here a social choice function $\psi :T\times S\rightarrow O$ specifies a feasible outcome for any pair $(t,s)$. 

A mechanism $\Gamma=(M,\gamma)$ consists of an arbitrary message space $M$ and a function $\gamma:M\times S\rightarrow O$. We assume any mixed strategies are already contained in the abstract set $M$. If the agent participates in the mechanism and sends a message $m\in M$, the mechanism specifies a state-dependent outcome $\gamma(m,s)$. In a direct mechanism the message space is restricted to be the type space $T$. A mechanism induces a decision problem for the agent in which she behaves in a particular way. In our setting, this behavior is given by a strengthening of the notion of optimality: a robust strategy must be better than any other message a type could send no matter what beliefs are considered.

\begin{definition}
Given a mechanism $\Gamma=(M,\gamma)$, a strategy $m :T\rightarrow M$ is \emph{robust in $\Gamma$} if for each $m \in M$,
\begin{equation*}
E_{\pi }[u(\gamma(m(t),s),t,s)] \geq E_{\pi }[ u(\gamma(m,s),t,s)] \ \ \ \forall \pi \in \Pi (t) 
\end{equation*}
\end{definition}

\noindent A mechanism robustly implements a particular social choice function when there exists a robust strategy in the mechanism that yields the outcome given by the social choice function for each type and state.

\begin{definition}
A social choice function $\psi $ is \emph{robustly implementable} by the mechanism $\Gamma$ if there exists a strategy  $m :T\rightarrow M $ that is robust in $\Gamma$ such that $\gamma(m(t),s) =\psi(t,s) $ for all $t\in T$ and for all $s\in S$.
\end{definition}

\noindent A social choice function is robustly truthfully implementable by a direct mechanism if the strategy of reporting the true type is a robust strategy in the direct mechanism, and the outcome of the direct mechanism is the same as the social choice function for each type and state.

\begin{definition}
A social choice function $\psi $ is \emph{robustly truthfully implementable} by the direct mechanism $\Gamma=(T,\gamma)$ if the function $m:T\to T$ such that $m(t)=t$ for each $t$ is a robust strategy in $\Gamma$, and $\gamma(t,s) =\psi(t,s)$ for each $t\in T$ and all $s\in S$.
\end{definition}

\noindent If $\psi $ is robustly truthfully implementable by the direct mechanism $\Gamma = (T,\gamma)$, then $\psi(t,s) = \gamma(t,s)$ for each $(t,s)$, and for each $t,t'\in T$, 
\begin{equation*}
E_{\pi }[u(\gamma(t,s),t,s)] \geq E_{\pi }[u(\gamma(t',s),t,s)] \ \ \ \forall \pi \in \Pi ( t) 
\end{equation*}

With these formalities in place, we can state the revelation principle for this setting.

\begin{proposition}[The Revelation Principle]
If a social choice function can be robustly implemented by some mechanism, then it can also be robustly truthfully implemented in a direct mechanism.
\end{proposition}

\begin{proof}
Let $\Gamma=(M,\gamma)$ be a mechanism that robustly implements the social choice function $\psi$, and let $m:T\to M$ be a corresponding robust strategy.
Since $\Gamma$ implements $\psi$, $\gamma(m(t), s) = \psi(t,s)$ for each $t$ and $s$. Then define the direct mechanism $\Gamma'=(T,\gamma')$ by setting 
$\gamma'(t,s)=\gamma(m(t),s) = \psi(t,s)$ for each $t$ and $s$. 

Then for each $t, t'\in T$, 
\begin{eqnarray*}
E_{\pi }[u(\gamma'(t,s),t,s)] &=&E_{\pi }[u(\gamma(m(t),s),t,s)]\\
&\geq&E_\pi [u(\gamma(m(t'),s), t, s)] = E_\pi(u(\gamma'(t',s), t,s) \ \ \ \forall \pi \in \Pi (t) 
\end{eqnarray*}
where the inequality follows from the fact that $m$ is a robust strategy in $\Gamma$. Thus the truthtelling strategy $m^*:T\to T$ with $m^*(t) = t$ for each $t$ is a robust strategy in $\Gamma'$, and $\Gamma'$ implements $\psi$. 
\end{proof}

\vfill\eject

\bibliographystyle{econ}
\bibliography{biblioluca}

@BOOK{Clarke,
  author = {Clarke, Frank H.},
  title = {Optimization and Nonsmooth Analysis},
  year = 1983,
  publisher = {Wiley},
  address = {New York}
}

@ARTICLE{Ioffe-Levin,
  author = {Ioffe, Aleksandr and Levin, Vladimir},
  title = {Subdifferentials of Convex Functions},
  journal = {Transactions of the Moscow Mathematical Society},
  year = {1972},
  volume = {26},
  pages = {3-73}
}

@ARTICLE{Borwein,
  author = {Borwein, Jonathan},
  title = {A Note on $\varepsilon$-subgradients and Maximal Monotonicity},
  journal = {Pacific Journal of Mathematics},
  year = {1982},
  volume = {103},
  pages = {307-314}
}

@ARTICLE{Ekeland,
  author = {Ekeland, Ivar},
  title = {On the Variational Principle},
  journal = {Journal of Mathematical Analysis and Applications},
  year = {1974},
  volume = {47},
  pages = {324-353}
}

@inproceedings{Debreu,
address = "Berkeley, Calif.",
author = "Debreu, Gerard",
booktitle = "Proceedings of the Fifth Berkeley Symposium on Mathematical Statistics and Probability, Volume 2: Contributions to Probability Theory,  Part 1",
pages = "351--372",
publisher = "University of California Press",
title = "Integration of Correspondences",
year = "1967"
}

@ARTICLE{Byrne,
  author = {Byrne, Charles},
  title = {Remarks on the Set-Valued Integrals of Debreu and Aumann},
  journal = {Journal of Mathematical Analysis and Applications},
  year = {1978},
  volume = {62},
  pages = {243-246}
}

@ARTICLE{Lopomo-Rigotti-Shannon-finite,
  author = {Giuseppe Lopomo and Luca Rigotti and Chris Shannon},
  title = {Uncertainty and Robustness of Surplus Extraction},
  journal = {Journal of Economic Theory},
  year = {2020},
  note = {forthcoming},
}

@BOOK{Zalinescu-02,
  author = {Zalinescu, Constantin},
  title = {Convex Analysis in General Vector Spaces},
  publisher = {World Scientific},
  year = {2002},
}

@unpublished{Lopomo-Rigotti-Shannon-detect,
    author  = "Giuseppe Lopomo and Luca Rigotti and Chris Shannon",
    title   = "Detectability, Duality, and Surplus Extraction",
    year    = "2020",
    note    = "working paper",
}

@ARTICLE{Kocyigit-Iyengar-Kuhn-Wiesemann-20,
  author = {Koçyiğit, Çağıl and Iyengar, Garud  and Kuhn, Daniel and Wiesemann, Wolfram},
  title = {Distributionally Robust Mechanism Design},
  journal = {Management Science},
  year = {2020},
  volume = {66},
  pages = {159-189},
  number = {1}
}

@ARTICLE{Ollar-Penta-17,
  author = {Ollár, Mariann and Penta, Antonio},
  title = {Full Implementation and Belief Restrictions},
  journal = {American Economic Review},
  year = {2017},
  volume = {107},
  pages = {2243-2277},
  number = {8}
}

@ARTICLE{Danan-Gajdos-Hill-Tallon,
  author = {Danan, Eric and Gajdos, Thibault and Hill, Brian and Tallon, Jean-Marc},
  title = {Robust Social Decisions},
  journal = {American Economic Review},
  year = {2016},
  volume = {106},
  pages = {2407-2425},
  number = {9}
}

@ARTICLE{Auster,
  author = {Auster, Sarah},
  title = {Robust Contracting Under Common Value Uncertainty},
  journal = {Theoretical Economics},
  year = {2018},
  volume = {13},
  pages = {175-204},
  number = {1}
}

@ARTICLE{Bergemann-Schlag-a,
  author = {Bergemann, Dirk and Schlag, Karl},
  title = {Pricing Without Priors},
  journal = {Journal of the European Economic Association},
  year = {2008},
  volume = {6},
  pages = {560-569},
  number = {}
}

@ARTICLE{Bergemann-Schlag-b,
  author = {Bergemann, Dirk and Schlag, Karl},
  title = {Robust Monopoly Pricing},
  journal = {Journal of Economic Theory},
  year = {2011},
  volume = {146},
  pages = {2527-2543},
  number = {}
}

@ARTICLE{Carrasco-Luz-Kos-Messner-Monteiro,
  author = {Carrasco, Vinicius and Luz, Vitor Farinha and Kos, Nenad and Messner, Matthias and Monteiro, Paulo and Moreira, Humberto},
  title = {Optimal Selling Mechanisms Under Moment Conditions},
  journal = {Journal of Economic Theory},
  year = {2018},
  volume = {177},
  pages = {245-279},
  number = {}
}

@unpublished{Suzdaltsev,
    author  = "Alex Suzdaltsev",
    title   = "An Optimal Distributionally Robust Auction",
    year    = "2020",
    note    = "working paper",
}

@unpublished{Hu-Haghpanah-Hartline-Kleinberg,
    author  = "Hu Fu and Nima Haghpanah and Jason Hartline and Robert Kleinberg",
    title   = "Full Surplus Extraction From Samples",
    year    = "2017",
    note    = "working paper",
}

@ARTICLE{Neeman,
  author = {Neeman, Zvika},
  title = {The Effectiveness of English Auctions},
  journal = {Games and Economic Behavior},
  year = {2003},
  volume = {43},
  pages = {214-238},
  number = {}
}

@ARTICLE{Alon-Gayer,
  author = {Alon, Shiri and Gayer, Gabi},
  title = {Utilitarian Preferences with Multiple Priors},
  journal = {Econometrica},
  year = {2016},
  volume = {84},
  pages = {1181-1201},
  number = {3}
}

@ARTICLE{Brunnermeier-Simsek-Xiong,
  author = {Brunnermeier, Marcus and Simsek, Alp and Xiong, Wei},
  title = {A Welfare Criterion for Models with Distorted Beliefs},
  journal = {Quarterly Journal of Economics},
  year = {2014},
  volume = {129},
  pages = {1753-1797},
  number = {4}
}

@ARTICLE{Gilboa-Samuelson-Schmeidler,
  author = {Gilboa, Itzak and Samuelson, Larry and Schmeidler, David},
  title = {No-Betting-Pareto Dominance},
  journal = {Econometrica},
  year = {2014},
  volume = {82},
  pages = {1405-1442},
  number = {4}
}

@unpublished{Ollar-Penta-19,
  author = {Ollár, Mariann and Penta, Antonio},
  title = {Implementation via Transfers with Identical but Unknown Distributions},
  year = 2019,
  institution = {Barcelona Graduate School of Economics},
  series = {Working Paper},
  note = {working paper}
}

@ARTICLE{Bodoh-Creed-12,
  author = {Bodoh-Creed, Aaron},
  title = {Ambiguous Beliefs and Mechanism Design},
  journal = {Games and Economic Behavior},
  year = {2012},
  volume = {75},
  pages = {518-537},
  number = {2}
}

@ARTICLE{Bose-Daripa-09,
  author = {Bose, Subir and Daripa, Arup},
  title = {A Dynamic Mechanism and Surplus Extraction under Ambiguity},
  journal = {Journal of Economic Theory},
  year = {2009},
  volume = {144},
  pages = {2084-2114},
  number = {5}
}

@ARTICLE{Bose-Renou-14,
  author = {Bose, Subir and Renou, Ludovic},
  title = {Mechanism Design with Ambiguous Communication Devices},
  journal = {Econometrica},
  year = {2014},
  volume = {82},
  pages = {1853-1872},
  number = {5}
}

@ARTICLE{DeTillio-Kos-Messner-17,
  author = {De Tillio, Alfredo and Kos, Nenad and Messner, Matthias},
  title = {The Design of Ambiguous Mechanisms},
  journal = {The Review of Economic Studies},
  year = {2017},
  volume = {84},
  pages = {237-276},
  number = {1}
}

@ARTICLE{Jehiel-Meyer-ter-Vehn-Moldovanu-12,
  author = {Jehiel, Philippe and Meyer-ter-Vehn, Moritz and Moldovanu, Benny},
  title = {Locally Robust Implementation and its Limits},
  journal = {Journal of Economic Theory},
  year = {2012},
  volume = {147},
  pages = {2439-2452},
  number = {6}
}

@ARTICLE{Chiesa-Micali-Zhu-15,
  author = {Chiesa, Alessandro and Micali, Silvio and Zhu, Zeyuan Allen},
  title = {Knightian Analysis of the Vickrey Mechanism},
  journal = {Econometrica},
  year = {2015},
  volume = {83},
  pages = {1727-1754},
  number = {5}
}

@ARTICLE{Wolitzky-16,
  author = {Wolitzky, Alexander},
  title = {Mechanism Design with Maxmin Agents: Theory and an Application to Bilateral Trade},
  journal = {Theoretical Economics},
  year = {2016},
  volume = {11},
  pages = {971-1004},
  number = {3}
}

@ARTICLE{Gilboa-Maccheroni-Marinacci-Schmeidler,
  author = {Gilboa, Itzhak and Maccheroni, Fabio and Marinacci, Massimo and Schmeidler, David},
  title = {Objective and Subjective Rationality in a Multiple Prior Model},
  journal = {Econometrica},
  year = {2010},
  volume = {78},
  pages = {755-770},
  number = {2}
}

@ARTICLE{Mertens-Zamir,
  author = {Mertens, Jean-François and Zamir, Shmuel},
  title = {Formulation of Bayesian Analysis for Games with Incomplete Information},
  journal = {International Journal of Game Theory},
  year = {1985},
  volume = {14},
  pages = {1-29}
}

@ARTICLE{Ahn07,
  author = {David Ahn},
  title = {Hierarchies of Ambiguous Beliefs},
  year = 2007,
  journal = {Journal of Economic Theory},
  volume  = 136,
  pages = {286-301}
}

@ARTICLE{Bergemann-Morris05,
  author = {Dirk Bergemann and Stephen Morris},
  title = {Robust Mechanism Design},
  year = 2005,
  journal = {Econometrica},
  volume = 73,
  pages = {1771-1813}
}

@ARTICLE{Bewley02,
  author = {Truman F. Bewley},
  title = {Knightian Decision Theory: Part I},
  year = 2002,
  journal = {Decisions in Economics and Finance},
  volume = 2,
  pages = {79-110}
}

@ARTICLE{Biketal06,
  author = {Sushil Bikhchandani and Shurojit Chatterji and Ron Lavi and Ahuva Mu'alem and Noam Nisan and Arunava Sen},
  title = {Weak Monotonicity Characterizes Deterministic Dominant-Strategy Implementation},
  year = 2006,
  journal = {Econometrica},
  volume = 74,
  pages = {1109-1132}
}

@ARTICLE{Bose-Ozdenoren-Pape-06,
  author = {Subir Bose and Emre Ozdenoren  and Andreas Pape},
  title = {Optimal Auctions with Ambiguity},
  year = 2006,
  journal = {Theoretical Economics},
  volume = 1,
  pages = {411-438}
}

@ARTICLE{CremerMcLean85,
  author = {Jean-Jaques Cr\'{e}mer and Richard McLean},
  title = {Optimal Selling Strategies under Uncertainty for a Discriminatory Monopolist when Demands Are Interdependent},
  year = 1985,
  journal = {Econometrica},
  volume = 53,
  pages = {345-61}
}

@ARTICLE{CremerMcLean88,
  author = {Jean-Jaques Cr\'{e}mer and Richard McLean},
  title = {Full Extraction of the Surplus in Bayesian and Dominant Strategy Auctions},
  year = 1988,
  journal = {Econometrica},
  volume = 56,
  pages = {1247-57}
}

@ARTICLE{Ellsberg,
  author = {Daniel Ellsberg},
  title = {Risk, Ambiguity, and the Savage Axioms},
  year = 1961,
  journal = {Quarterly Journal of Economics},
  volume = 75,
  pages = {643-669}
}

@ARTICLE{GMMS,
  author = {Ghirardato, Paolo and Maccheroni, Fabio and Marinacci, Massimo and Siniscalchi, Marciano},
  title = {A Subjective Spin on Roulette Wheels},
  journal = {Econometrica},
  year = {2003},
  volume = {71},
  pages = {1897 - 1908},
  number = {6}
}

@ARTICLE{Ghirardato-Maccheroni-Marinacci,
  author = {Ghirardato, Paolo and Maccheroni, Fabio and Marinacci, Massimo},
  title = {Differentiating Ambiguity and Ambiguity Attitude},
  year = {2004},
  journal = {Journal of Economic Theory},
  volume = {118},
  pages = {133-173},	
  number = {2}
}

@BOOK{Knight21,
  author = {Frank H. Knight},
  title = {Uncertainty and Profit},
  year = 1921,
  publisher = {Houghton Mifflin},
  address = {Boston}
}

@BOOK{Krishna02,
  author = {Vijay Krishna},
  title = {Auction Theory},
  year = 2002,
  publisher = {Academic Press},
  address = {}
}

@ARTICLE{Ledyard78,
  author = {John O. Ledyard},
  title = {Incentive Compatibility and Incomplete Information},
  year = 1978,
  journal = {Journal of Economic Theory},
  volume = 18,
  pages = {171-189}
}

@INCOLLECTION{Ledyard79,
  author = {John O. Ledyard},
  editor = {Jean-Jacques Laffont},
  title = {Dominant Strategy Mechanisms and Incomplete Information},
  year = 1979,
  volume = {},
  booktitle = {Aggregation and Revelation of Preferences},
  publisher = {North-Holland},
  address = {Amsterdam},
 }

@ARTICLE{McAfeeReny92,
  author = {Preston R. McAfee and Philip J. Reny},
  title = {Correlated Information and Mechanism Design},
  year = 1992,
  journal = {Econometrica},
  volume = 60,
  pages = {395-421}
}

@ARTICLE{Rigotti-Shannon-Strzalecki08,
  author = {Luca Rigotti and Chris Shannon and Tomasz Strzalecki},
  title = {Subjective Beliefs and ex ante Trade},
  year = {2008},
  journal = {Econometrica},
  volume = 76,
  issue = 5,
  pages = {1167-1190}
}

@ARTICLE{Rochet87,
  author = {Jean-Charles Rochet},
  title = {A Necessary and Sufficient Condition for Rationalizability in a Quasi-linear Context},
  year = 1987,
  journal = {Journal of Mathematical Economics},
  volume = 16,
  issue  = 2,
  pages = {191-200}
}

@ARTICLE{Rigotti-Shannon,
  author = {Luca Rigotti and Chris Shannon},
  title = {Uncertainty and Risk in Financial Markets},
  year = 2005,
  journal = {Econometrica},
  volume = 73,
  pages = {203-243}
}

\end{document}